\documentclass[12pt]{iopart}

\usepackage{graphicx}
\usepackage{dcolumn}
\usepackage{epstopdf}
\usepackage[usenames]{color}
\usepackage{color}
\usepackage{amssymb}
\usepackage{bm}
\usepackage{hyperref}
\usepackage{lineno}
\usepackage{cite}

\graphicspath{{figure/}}

\newcommand{\pT}{$p_{\rm{T}}$}

\newcommand{\sNN}{$\sqrt{s_{\rm NN}}$}

\newcommand{\AonA}{A--A }
\newcommand{\pp}{p--p}
\newcommand{\ppbar}{p--$\rm \bar p$}
\newcommand{\pA}{p--A}

\newcommand{\dAu}{d--Au}
\newcommand{\pPb}{p--Pb}

\newcommand{\AuAu}{Au--Au}
\newcommand{\PbPb}{Pb--Pb}
\newcommand{\CuCu}{Cu--Cu}

\newcommand{\la}{\langle}
\newcommand{\ra}{\rangle}

\begin{document}
\title[Multiplicity and pseudorapidity density
distributions]{Multiplicity and pseudorapidity density distributions
  of charged 
particles produced in pp, pA and AA collisions at RHIC \& LHC energies}

\author{Sumit Basu$^{1,2}$,  Sanchari Thakur$^3$, Tapan K. Nayak$^{4,5}$, and Claude A. Pruneau$^1$}

\address{$^1$ Department of Physics \& Astronomy, Wayne State University, MI 48201 USA \\
 $^2$ Lund University, Department of Physics, Division of Particle Physics, Box 118, SE-221 00, Lund, Sweden \\
 $^3$ Variable Energy Cyclotron Centre, HBNI, Kolkata-700064, India \\
 $^4$ National Institute of Science Education and Research, HBNI, Jatni 752050, India \\
 $^5$ CERN, CH 1211, Geneva 23, Switzerland}

 \ead{sumit.basu@cern.ch}
 \ead{sanchari.thakur@cern.ch}
 \ead{tapan.nayak@cern.ch}
 \ead{claude.pruneau@wayne.edu}
 \date{ \today }

\begin{abstract}

Multiplicity and pseudorapidity ($\eta$) density ($dN_{\rm ch}/d\eta$) 
distributions of charged hadrons provide key information towards 
understanding the particle production mechanisms and initial conditions 
of high-energy heavy-ion collisions. 
However, detector constraints limit the $\eta$-range across which  
charged particle measurements can 
be carried out. Extrapolating the measured distributions to large $\eta$-range
by parameterizing  measured distributions and by
using calculations from  event generators, we 
characterize the production of charged particles over the full
kinematic range. In the present study, we use 
three different ans$\ddot{\mathrm a}$tze
to obtain quantitative descriptions of  
the shape  of pseudorapidity distributions of charged hadrons produced in pp, p-A, and A-A collisions 
for beam energies ($\sqrt{s_{\rm NN}}$) ranging from a few GeV to a few TeV corresponding to RHIC and LHC energies.
We study the limiting fragmentation behavior in these collisions and 
report evidence for participant-scaling violations in high-energy collisions at 
the TeV scale. We additionally examine measured pseudorapidity distributions to 
constrain  models describing initial conditions of particle 
production. We predict the centrality dependence of charged particle multiplicity distributions at FAIR and NICA energies and give an estimation of charged particle multiplicity at $\eta=0$ for the proposed HE-LHC and FCC energies.

\end{abstract}

\maketitle
\section{Introduction}

Multiplicity and pseudorapidity ($\eta$) density distributions of
charged particles along with the
transverse momentum (\pT) spectra constitute some of the basic
observables for understanding the particle production mechanisms 
in high-energy elementary particle and heavy-ion collisions~\cite{bialas,bjorken1,kharzeev,jan}. 
The dependence of these distributions on the colliding particle
species, collision energy, and collision centrality have been
extensively discussed in the literature~\cite{Phobos_data_all,star_bulk,alice_bulk,raghu1,raghu2,Sumit}.
In  proton-proton (pp) collisions, 
these distributions provide precise calibration of particle production models
such as PYTHIA~\cite{pythia1} and HERWIG~\cite{herwig}, which are used to make predictions of various
searches including those of physics beyond the 
standard model. 
These measurements play an important role in the study of heavy-ion 
collisions at ultra-relativistic 
energies in which short-lived systems consisting of 
nuclear matter at extreme conditions of temperature and energy 
density are created. There is evidence that this matter undergoes a phase transition from a confined 
state to a de-confined state of quark-gluon plasma
(QGP)~\cite{star_white,heinz}.
The comparison of the charged particle distributions in 
pp, \pA, and \AonA\ collisions are essential to
characterize the formation of QGP and understand the particle production 
mechanisms.

The measured charged-particle multiplicity and \pT\ distributions are
dominated by  final state interactions and the state of matter at
freeze-out. Nonetheless, these distributions are also sensitive to
the initial stages of the collision. 
At small Bjorken-$x$ (expressed as $x=\frac{p_{\rm T}}{\sqrt{s}} . e^{-y} \sim \frac{p_{\rm
T}}{\sqrt{s}} . e^{-\eta}$, $y$ being the rapidity),
the gluon density of the parton distribution functions (PDF) of the
proton grows and is expected  to reach 
a saturation domain~\cite{Larry_saturation,Larry1,Larry2}. 
So the particle productions at large collision
energies and forward rapidities are characterized 
by a large number of gluons ~\cite{cgc1,cgc2,cgc3,duraes,lappi, rezaeian}. 
With present theoretical understandings and available deep inelastic scattering (DIS) experiments at HERA~\cite{ZEUS,H1}, it becomes possible to study the
expected growth and saturation of gluon density at high collision energies.  
The measurements of produced charged-particle multiplicity,
$p_{\rm T}$, and $\eta$ distributions are typically restricted to the mid-rapidity
 region. Such restrictions arise in part because of   favorable
kinematic conditions at mid-rapidity and largely because of experimental limitations at forward rapidities. 
Understanding the particle production dynamics, including
effects of nuclear stopping, color transparency, jet quenching, and
long range correlations, requires the measurement of particle
production in full pseudorapidity ranges. 
However, energy-momentum conservation dictates that particle production must vanish at or beyond the
beam rapidity. It is thus of interest to consider pseudorapidity ansatz that assumes vanishing density 
at such large rapidities. These may then be compared to production models and may, in principle, 
be used to estimate the total charged-particle production in pp and \AonA\  collisions.  The 
precision achievable with such extrapolations is obviously limited by the quality of the ansatz but it can 
be tested with existing pp and \AonA\  collision models. 
However, no specific or widely accepted pseudorapidity rapidity model
is currently available  in the literature to carry out
such extrapolation. In this work, we exploit 
the large body of  available experimental data measured in high-energy  pp,  \pA, and \AonA\  collisions 
to examine and compare the merits of three ans$\ddot{\mathrm a}$tze towards a phenomenological 
description of  pseudorapidity density as well as the extrapolation and integration of 
measured densities to estimate total charged-particle production with beam energy.  Our analysis is 
based on data collected from a variety of collision systems 
and for collision energies ranging from a few GeV to the top LHC energy.
These distributions, at close to beam rapidities, are used  to
study the limiting fragmentation of particle production~\cite{benecke,raha,gelis}.

Total particle multiplicities and pseudorapidity densities at
mid-rapidity at CERN SPS (Super Proton Synchrotron) and
RHIC (Relativistic Heavy Ion Collider) energies have been observed to
be proportional to the number of participating nucleons  ($N_{\rm part}$)~\cite{npart1,npart2}. 
But at higher energies of the CERN Large Hadron Collider (LHC), the
$N_{\rm part}$ scaling has been observed to be
broken~\cite{ALICE_XeXe,ALICE_PbPb}. 
One of the main reasons for this scale breaking is the 
enhancement in gluon productions at high energies
(low Bjorken-$x$)~\cite{Larry1,Larry2,Larry3}.

The paper is organised as follows.
In sec.~\ref{sec:expData}, we examine the measured charged-particle
multiplicity, and pseudorapidity distributions, $dN_{\rm ch}/d\eta$, observed in pp, \pA, and \AonA\ collisions
across a wide range of beam energies and  compare these  with results from selected event generators. 
In sec.~\ref{sec:Param}, we parameterize these $dN_{\rm ch}/d\eta$
distributions using three different ans$\ddot{\mathrm a}$tze to obtain a satisfactory 
model one can integrate over the full pseudorapidity range spanned by particle production. 
Such an extrapolation requires that we examine, in sec.~\ref{sec:limitingFragmentation},  the measured distributions
in the vicinity of the beam rapidity and study the applicability  of the notion of limiting fragmentation.
In sec.~\ref{sec:TotalMult}, we use the favored ansatz to estimate the total number of charged
particles produced per $N_{\rm part}$ as a function of collision energy and
centrality. We inspect whether the charged particle production
 scales with $N_{\rm part}$ irrespective of the collision energy. Using the parameterization of the pseudorapidity density distributions, we give  predictions for these distributions as well as total charged particle multiplicities for lower collision energies of the future experiments at FAIR (Facility for Anti-proton  and Ion Research) at GSI, Germany and NICA (Nuclotron-based Ion Collider fAcility) at JINR, Russia.
 Additionally, we extend the charged particle multiplicity density at $\eta=0$ for the proposed HE-LHC (High-Energy LHC) and FCC (Future Circular Collider) at CERN.
 Finally, in sec.~\ref{sec:InitialConditions}, we explore whether selected  initial condition scenarios can be meaningfully constrained by measured particle multiplicity distributions. The paper is summarized in sec.~\ref{sec:summary}.

\section{Charged-particle multiplicity distributions}
\label{sec:expData}

In this section, we present the charged particle multiplicity density
at mid-rapidity and pseudorapidity distributions from available
experimental data for pp, \ppbar, \dAu, \pPb, \AuAu, and \PbPb \space collisions. These data are compared to calculations from event
generators.
For pp and \ppbar\ collisions, the multiplicities
are calculated with PYTHIA (Perugia tune)~\cite{pythia1}, whereas those for  \AonA\ collisions 
are computed with 
UrQMD~\cite{UrQMDapply1,UrQMDapply2,Bleichersus,Arghya},
AMPT~\cite{ampt1,ampt2,ampt3}, EPOS (we have used EPOS-LHC v3.4) ~\cite{EPOS-param,EPOS3,EPOS4}, and 
THERMINATOR~\cite{thermi}. UrQMD (Ultrarelativistic Quantum 
Molecular Dynamics) is a microscopic transport model based 
on the covariant propagation of all produced hadrons in combination with 
stochastic binary scatterings, color string formation, and 
resonance decay. It has been widely 
used to simulate the production of different particles, particle flow and 
fluctuations. 
AMPT (A Multi-Phase Transport) models the initial stage of \AonA\ collisions 
in terms of  partonic 
interactions. It converts produced partons into hadrons and includes a 
hadronic interactions stage~\cite{ampt1,ampt2,ampt3}. 
AMPT calculations have been carried out with the string melting (SM) option, 
which involves a fully partonic QGP phase that 
hadronizes through quark coalescence. 
EPOS 
 is 
 a hybrid event generator describing \AonA\ as well as pp collisions in terms of a core (high density) and corona (low density) components~\cite{EPOS-param,EPOS4}. 
It describes the evolution of the core component   with a viscous hydrodynamical model 
while collisions within the corona are computed with Gribov-Regge (GR) theory and 
perturbative QCD~\cite{EPOS-param,EPOS3}. The core/corona approach is known to successfully 
reproduce the measured collision centrality evolution of several observables, including relative 
particle abundance ratios, transverse $p_{\rm T}$ distributions, and 
anisotropic flow~\cite{EPOS1,EPOS-param,EPOS3,EPOS4,EPOS5}. 
THERMINATOR (THERMal heavy IoN generator) is a statistical hadronization 
model commonly used to estimate the relative abundances of particles species 
produced  in relativistic heavy-ion collisions. It enables arbitrary 
implementations of the  shape 
of the freeze-out hyper surface and the expansion velocity field. 
The multiplicities were computed including HBT effects and 
3+1 dimensional profiles~\cite{thermi}. 

\subsection{Charged-particle multiplicity density at mid-rapidity}

\begin{figure}[!th]
\centering \includegraphics[width=0.8\textwidth]{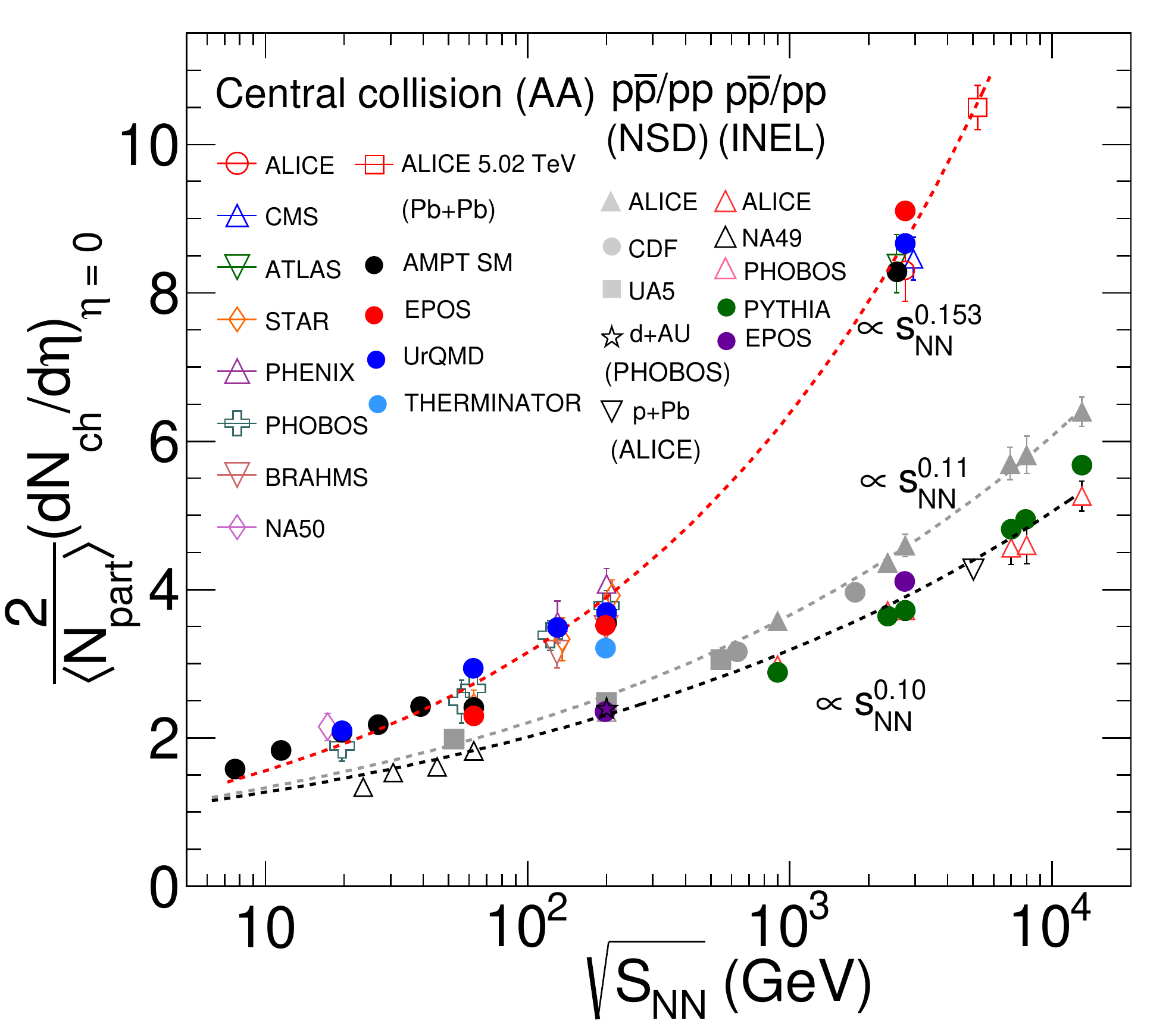}
\caption{Compilation of measurements of the beam-energy dependence of charged-particle multiplicity 
  density at mid-rapidity, scaled by the average number of participating nucleon 
  pair ($\la N_{\rm part}\ra/2$). Data from  pp, \ppbar, \dAu, \pPb,
  \AuAu, and \PbPb~ collisions are parameterized with power-law fits (dash-lines) and  compared to calculations from event generators. }
\label{fig:model1}
\end{figure}

The charged-particle multiplicity density at mid-rapidity 
$\left. dN_{\rm ch}/d\eta  \right|_{\eta=0}$ has been reported for different
colliding systems, collision centrality and collision energies. 
The average number of participants ($\langle N_{\rm part}\rangle$)
characterizes the collision centrality and colliding system. 
In Fig.~\ref{fig:model1}, we present a compilation of the measurements 
 of scaled charged-particle multiplicity density at mid-rapidity,
 $\frac{2}{\langle N_{\rm part}\rangle} \left. dN_{\rm ch}/d\eta 
 \right|_{\eta=0}$,  
as a function of collision energy in  
pp~\cite{pp_900_2360,pp_2760_7_8,pp_13},  
\ppbar~\cite{UA5,pp_630_1800},  
\AuAu~\cite{Phobos_data_all,systematic_star,systematic_phobos,eta_200_19,eta_Au_130,eta_dist_Au_62,eta_dist_Cu}, 
\PbPb~\cite{mul_dens_PbPb5020,eta_dist_2760,Pb_5020_eta},
\dAu~\cite{eta_dAu}, and \pPb~\cite{pPb_eta_dist,eta_pPbs,Ncoll_pPb} collisions observed  at Fermilab, 
RHIC, and LHC energies.
Results from  pp and \ppbar\ collisions are  for non-single 
diffractive (NSD) as well as  inelastic (INEL) collisions, 
 whereas those from \AuAu\ and \PbPb\ collisions correspond to most central collisions. 

The multiplicity densities measured in  pp (\ppbar)  and \AonA 
collisions exhibit  rather different dependence as a function of
collision energy. 
These dependencies can be characterised with power-law fits performed separately for \AonA,  
NSD pp, and INEL pp (\ppbar) collisions.
We find that  the $\sqrt{s}$ dependence of the  multiplicity density
of  pp  (\ppbar) collisions  are well 
matched by power laws of  the 
 form $(s_{\rm NN})^{\alpha}$ with exponent $\alpha=0.10$ and  $\alpha=0.11$ for INEL and NSD collisions,
 respectively. In contrast, the multiplicity densities observed in \AonA\ collisions exhibit a steeper dependence on the beam energy 
 and are best described with a power law exponent $\alpha=0.153$.  Additionally, we find that the dependences of the   multiplicity 
densities achieved in \dAu\ and \pPb\ collisions are similar to those observed in  pp collisions.

Comparing the results from different models and data shown in Fig.~\ref{fig:model1}, we note that  
for the \pp\ collision system, PYTHIA predictions are in good
agreement with INEL data for beam energies $\sqrt{s_{\rm NN}}\geq 100$ GeV. In the case of  \AonA\ systems, one finds that AMPT SM and UrQMD predictions are in 
 good agreement with data over the entire  $\sqrt{s_{\rm NN}}$ range considered in this work. We additionally find that
 EPOS predictions are also in reasonable agreement with data from both \pp\  and \AonA\ systems over a wide range of 
beam energies. However, the single  THERMINATOR prediction considered at $\sqrt{s_{\rm NN}}=200$ GeV   is found to considerably underestimate the measured charged-particle density. Overall,  PYTHIA , EPOS, AMPT, and UrQMD 
are found to reproduce reasonably well  the observed  \sNN\  power law behavior even though they are based on 
rather different interaction and transport models. The \sNN\  evolution of the $\frac{2}{\langle N_{\rm part}\rangle} \left. dN_{\rm ch}/d\eta \right|_{\eta=0}$, an integrated observable, is  not a strong discriminant of these models and their underlying particle production mechanisms. 
Indeed, although historically  important,  measurements of scaled charge particle density at central rapidity provide only a rather limited amount of information about the specific particle
creation and transport mechanisms involved in pp and \AonA\ 
collisions~\cite{Sumit}. Additional insight into these mechanisms, however, may be gained from charged-particle pseudorapidity distributions.  Figure~\ref{fig:model2}  presents a
compilation of    $\frac{1}{\sigma}d\sigma/d\eta$ distributions measured in pp, \AuAu, and \PbPb\  collisions at the  SPS, RHIC and LHC.  

\subsection{Pseudorapidity distributions}

\begin{figure}[tbp]
	\centering \includegraphics[width=0.8\textwidth]{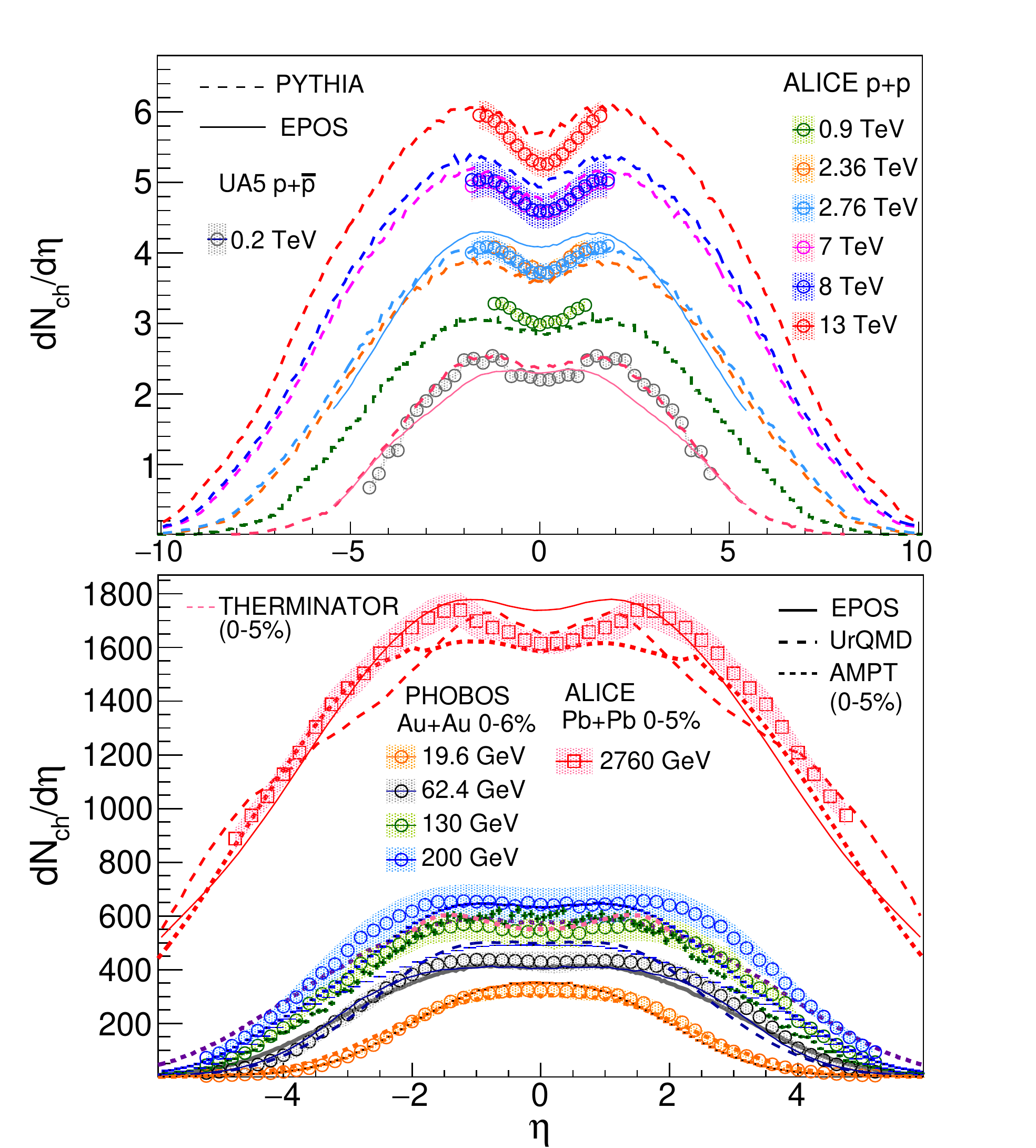}
	\caption{Comparison of selected experimental $dN_{\rm ch}/d\eta$ distributions of 
         measured in  pp and \ppbar~collisions  (upper panel) and 
          \AuAu\  and \PbPb\  collisions (lower panel) with calculations performed with 
          the PYTHIA, AMPT, UrQMD, EPOS and THERMINATOR models.}
\label{fig:model2}
\end{figure}

The upper panel of Fig.~\ref{fig:model2} displays   distributions measured in pp collisions at energies 
ranging from 0.9  to 13~TeV~\cite{pp_900_2360,pp_2760_7_8,pp_13,pp_630_1800}, and \ppbar\ collisions at 0.2~TeV~\cite{UA5} (open symbols), while the lower panel
presents pseudorapidity
distributions of charged hadrons measured in  6\% most central  \AuAu\ collisions in the range $19.6 \le \sqrt{s_{\rm NN}} \le 200$~GeV
~\cite{Phobos_data_all,eta_200_19,eta_Au_130,eta_dist_Au_62,eta_dist_Cu,eta_dAu},  
and top 5\% central \PbPb\ collisions at 2.76 TeV~\cite{mul_dens_PbPb5020,eta_dist_2760}.

First focusing our attention to the upper panel of Fig.~\ref{fig:model2}, we note that only the UA5 data 
cover  a wide enough pseudorapidity
range capable of revealing the full shape of the $\eta$ distribution measured in pp collisions while the measurements reported by the ALICE collaboration are limited  to  the central 
rapidity region in pp collisions. One nonetheless observes that the  particle density produced in pp (\ppbar)
collisions rises monotonically, as expected, with beam energy. One also notes that the measured pseudorapidity distributions all feature a dip, centered at  $\eta=0$, whose  depth and width  increases with rising $\sqrt{s}$. The presence of this dip is in part associated with partial 
transparency and limited stopping power of the colliding protons (or anti-protons)~\cite{wolschin}. The dip may also result, in part, from the measurement being reported as a function of pseudorapidity. A boost invariant rapidity distribution would indeed yield  a broad dip in pseudorapidity owing to the presence of mass term in the denominator of $y\rightarrow \eta$ transformation Jacobian. 

The  pseudorapidity distributions measured in pp (\ppbar)
collisions are compared with Monte Carlo calculations
performed with PYTHIA 6.4~\cite{pythia1} (dash lines) and   EPOS~\cite{EPOS1} (solid lines) event generators. One observes that both PYTHIA and
EPOS approximately reproduce the magnitude and $\eta$ dependence of
the distributions:  both models indeed capture the essential features
of the measured distributions, including the presence of the widening
and deepening dip, centered at  $\eta=0$, with increasing $\sqrt{s}$. However, PYTHIA appears to
be in better agreement with the data than EPOS  at $\sqrt{s}=$
0.2, 2.76, and 8 TeV. Observe, in particular, that EPOS does not reproduce the 
dip structure seen in \ppbar\ data at 0.2 TeV.

Let us next examine the pseudorapidity distributions reported by the PHOBOS 
and ALICE collaborations  shown in the lower panel 
of Fig~\ref{fig:model2}.  The  PHOBOS data 
cover the  range $-5.4 \le \eta \le 5.4$ whereas those of the ALICE experiment
span the range $-5.0 \le \eta \le 5.5$. One finds that the  pseudorapidity distribution observed 
at $\sqrt{s_{\rm NN}}$ = 19.6 GeV features an approximate Gaussian shape peaked at $\eta=0$, while distributions observed at higher beam-energy are much broader and feature  dip structures qualitatively similar to those observed in pp collisions. However,  the dip structures observed in \AonA\ collisions are typically shallower and wider than
those observed in pp collisions. 

The experimental data are compared to calculations based
on the UrQMD, AMPT, EPOS, and THERMINATOR models shown as solid lines in Fig.~\ref{fig:model2}.
The calculations were performed within  $p_{\rm T}$ ranges reproducing the experimental 
conditions of the PHOBOS and ALICE experiments. Overall, we note that all four models 
qualitatively reproduce the observed distributions. 
However, we also note that best agreement with the measured pseudorapidity density 
distributions is obtained with the EPOS model
at beam energies $\sqrt{s_{\rm NN}}\le 200$ GeV, while at 2.76 TeV, none 
of the models reproduce the observed distributions quantitatively. 
Overall, all four models considered manage to capture the general
trend of the observed data, 
including the produced particle density and its dependency on pseudorapidity, but 
none perfectly reproduce the  shape  of the measured distributions. EPOS arguably works very well given that it matches not only the amplitude and breadth of the distributions but it
also produces a dip near $\eta=0$, albeit with insufficient depth. EPOS' predictions are ~$\sim 5\%$ low at RHIC energies and approximately ~$\sim 2\%$ high at 2.76 TeV.   AMPT and UrQMD, on the other hand,  seem to reproduce the measured densities rather well at mid-rapidity, across the entire \sNN\ range presented in Fig.~\ref{fig:model2}, but fail to reproduce the overall shape at higher $\eta$ values.  THERMINATOR, on the other hand, is doing 
rather poorly at $\sqrt{s} = 200$ GeV.

\section{Parameterization of the pseudorapidity  distributions}
\label{sec:Param}

Although the  PHOBOS and ALICE data  presented in the lower panel of Fig.~\ref{fig:model2}
cover  large ranges of pseudorapidity, they do not  capture the full range of particle 
production involved at top RHIC energy and at the LHC. In fact,
most reported measurements of charged-particle pseudorapidity distributions 
are limited to somewhat  narrow  ranges of pseudorapidity and do not account for
the full particle production.  For instance, the measured distributions reported by the ALICE collaboration 
for Pb-Pb collisions cover a fairly wide range, $|\eta|<5.5$, 
but this range is quite narrow relative to the beam rapidity  ($y_{\rm beam} \sim~9.0$). One may then wonder how much particle production  actually takes place at forward/backward rapidities? 
Can  the stark  differences between the  $\sqrt{s_{\rm NN}}$ dependence 
of the multiplicity densities  observed in  pp and \AonA\ collisions
result  from an overall  increase in the produced multiplicity per
participant pair or does it result simply from  a shift in the particle production 
towards  central rapidity, due possibly to larger stopping in \AonA\ collisions?

Very few experiments
are equipped to cover the entire pseudorapidity range required to
answer these questions unambiguously. 
In the absence of  such measurements, we seek to 
parameterize the measured $\eta$ distributions to obtain sensible extrapolations at forward/backward rapidities that may be used to estimate the total charged-particle production.  

In the transverse direction, extrapolation of measured particle densities, $\frac{1}{p_{\rm  T}} \frac{dN}{dp_{\rm  T}}$, to zero and infinite $p_{\rm T}$ is relatively straightforward because
the cross-section must vanish at these limits~\cite{CMS_pp}.   Evidently, models used to integrate $p_{\rm T}$ spectra
are not constrained outside of the fiducial $p_{\rm T}$ acceptance,
but the fact that the cross-section vanishes at null
and infinite $p_{\rm T}$ significantly constrains the shape of $p_{\rm
  T}$ distributions. Uncertainties as to the exact shape of the $p_{\rm T}$
  distribution outside of the fiducial acceptance thus   yield systematic uncertainties
  on the integral of the distributions. 

We seek to  use  the same concept towards extrapolating at forward and backward 
rapidities beyond the fiducial pseudorapidity acceptance. The pseudorapidity density 
must obviously vanish at suitably large values of $|\eta|$ but 
extrapolation beyond the  measurement acceptance does
require one makes assumptions about the  overall shape of the distributions. In this work, we explore 
three fitting functions 
to fit  and extrapolate measured yields beyond  their fiducial ranges. These functions can then be 
integrated numerically  over the entire range of particle production to
obtain (extrapolated) total produced particle multiplicities.

\begin{figure}[!th]
	\centering	\includegraphics[width=1.0\textwidth]{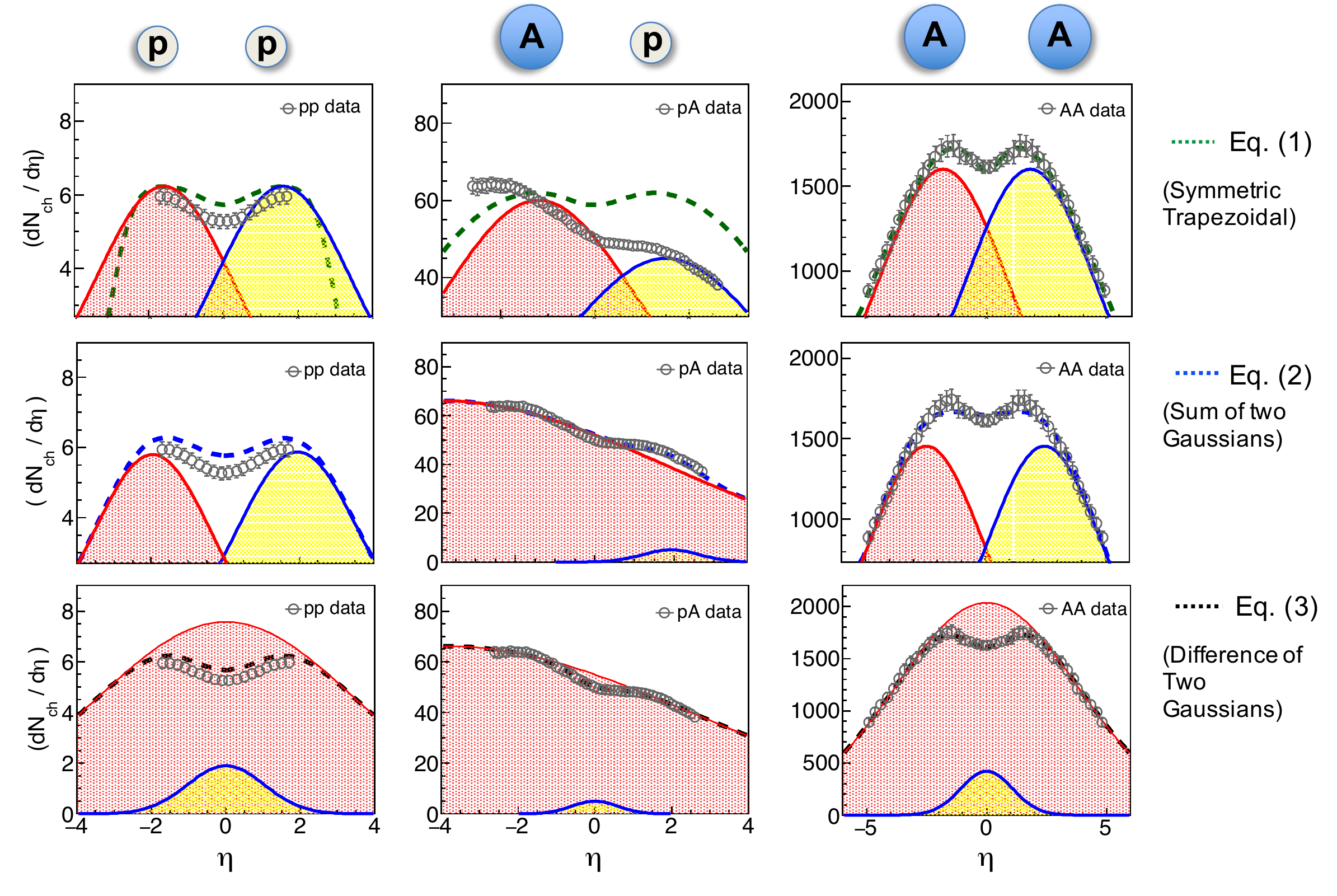} 
	\caption{Schematic representation of $dN_{\rm ch}/d\eta$ distributions of pp, \pA, and 
		\AonA\ collisions using three different ans$\ddot{\mathrm a}$tze.
		The distributions shown as dotted lines
		are parameterized with three fit functions: symmetic trapezoidal (upper 
		panels), sum of two Gaussian distributions (middle panels), and difference of 
		two Gaussian distributions (lower panels).}
	\label{fig:schematic}
\end{figure}

The analysis of the shapes of the pseudorapidity distributions is based on 
the  distributions  presented in Fig.~\ref{fig:model2}. 
We first note that the  pseudorapidity density distributions produced in symmetric collisions (e.g., pp and \AonA) are symmetric
about $\eta=0$, but distributions of the pA collisions feature a pronounced forward/backward asymmetry, with an
excess of particles being produced in the nucleus direction. 
We further note that the shape of the pseudorapidity distributions may 
be characterized by one broad peak with approximate Gaussian shape or 
two peaks of approximately Gaussian shape separated by a trough. 
For  simplicity, we thus consider three basic shapes defined according to: 
\begin{eqnarray}
\label{eq:fT}
f_T(\eta)&=&\frac{c\sqrt{1-1/(\alpha \cosh\eta)^2}}{1+e^{(|\eta|-\beta)/a}}, \\ 
\label{eq:fGPG}
f_{G+G}(\eta)&=&A_{1}e^{-\frac{(\eta-\mu_{1})^2}{2\sigma^2_{1}}}+A_{2}e^{-\frac{(\eta-\mu_{2})^2}{2\sigma^2_{2}}},  \\ 
\label{eq:fGMG}
f_{G-G}(\eta)&=&A_{1}e^{-\frac{\eta^2}{2\sigma^2_{1}}}-A_{2}e^{-\frac{(\eta-\mu)^2}{2\sigma^2_{2}}}.
\end{eqnarray}

\begin{figure*}[!th]
	\centering \includegraphics[width=1.0\textwidth]{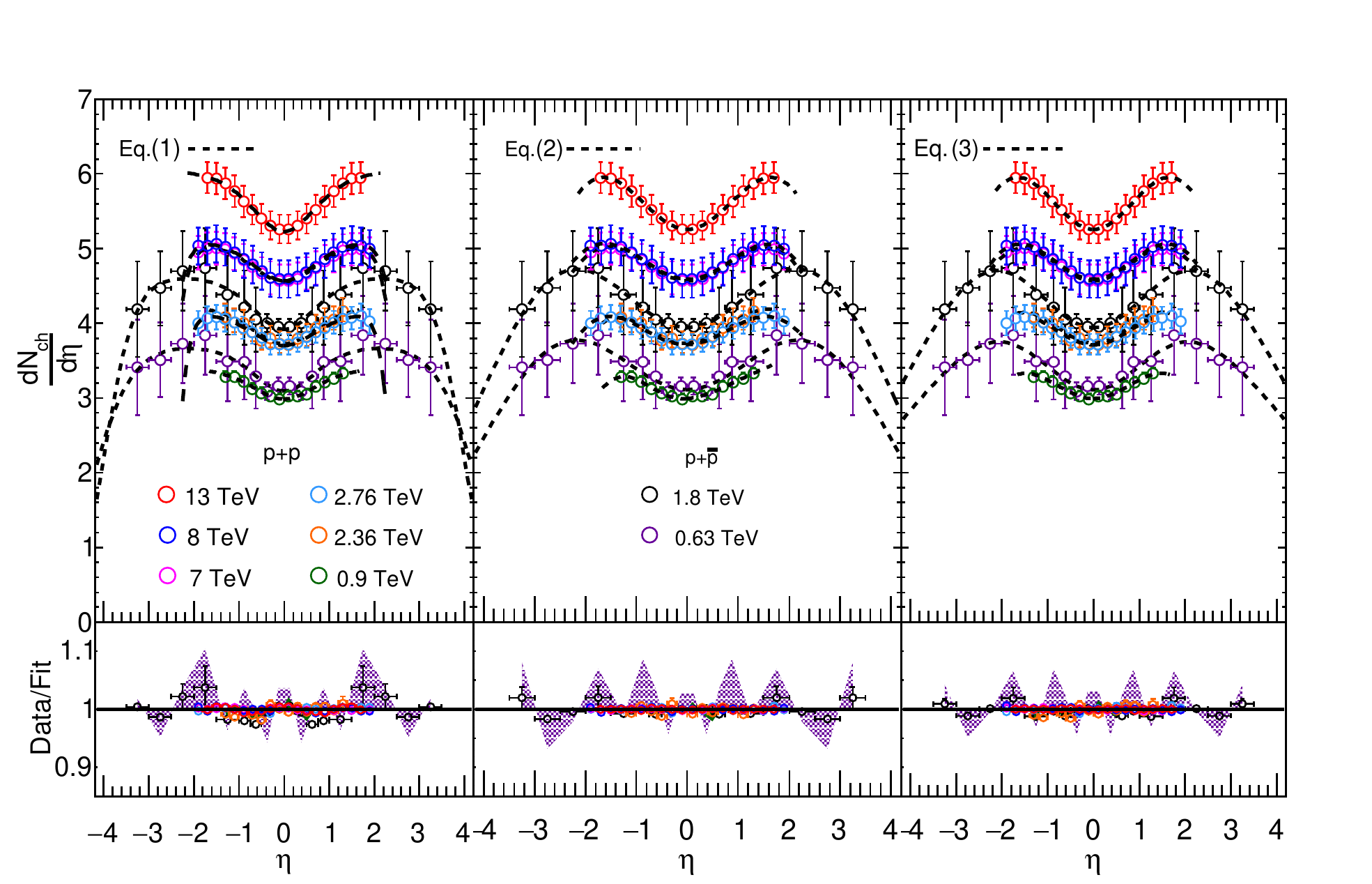}
	\caption{Beam energy dependence of charged 
		particle pseudorapidity density distributions measured in minimum bias pp 
		collisions by the ALICE collaboration~\cite{pp_900_2360,pp_2760_7_8,pp_13} and in \ppbar\ collisions by the CDF 
		collaboration~\cite{pp_630_1800}. Dashed lines show best fits obtained with Eqs.~(\ref{eq:fT}-\ref{eq:fGMG}).
	}
	\label{fig:etapp}
\end{figure*}

\begin{figure*}[!th]
	\centering \includegraphics[width=1.0\textwidth]{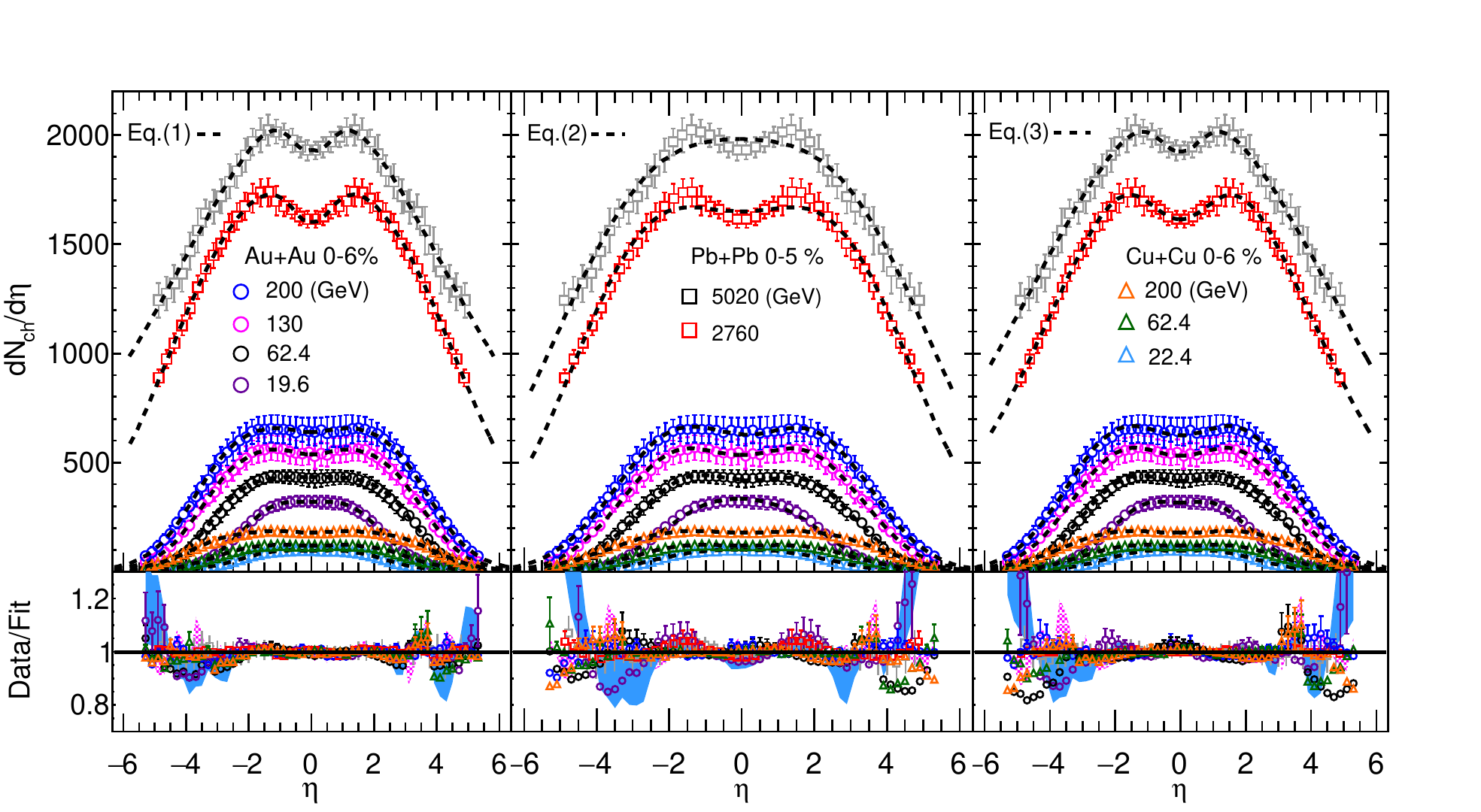}
	\caption{Beam energy dependence of charged 
          particle pseudorapidity density distributions measured in central \CuCu\ and \AuAu\  collisions by the 
          PHOBOS collaboration~\cite{eta_200_19,eta_Au_130,eta_dist_Au_62,eta_dist_Cu}
          and in \PbPb\ collisions by the  ALICE collaboration~\cite{eta_dist_2760}. 
          Dashed lines show best fits obtained with Eqs.~(\ref{eq:fT}-\ref{eq:fGMG}) from left to right panels respectively. 
}
\label{fig:etaAA}
\end{figure*}

\begin{figure*}[!th]
		\centering\includegraphics[width=1.0\textwidth]{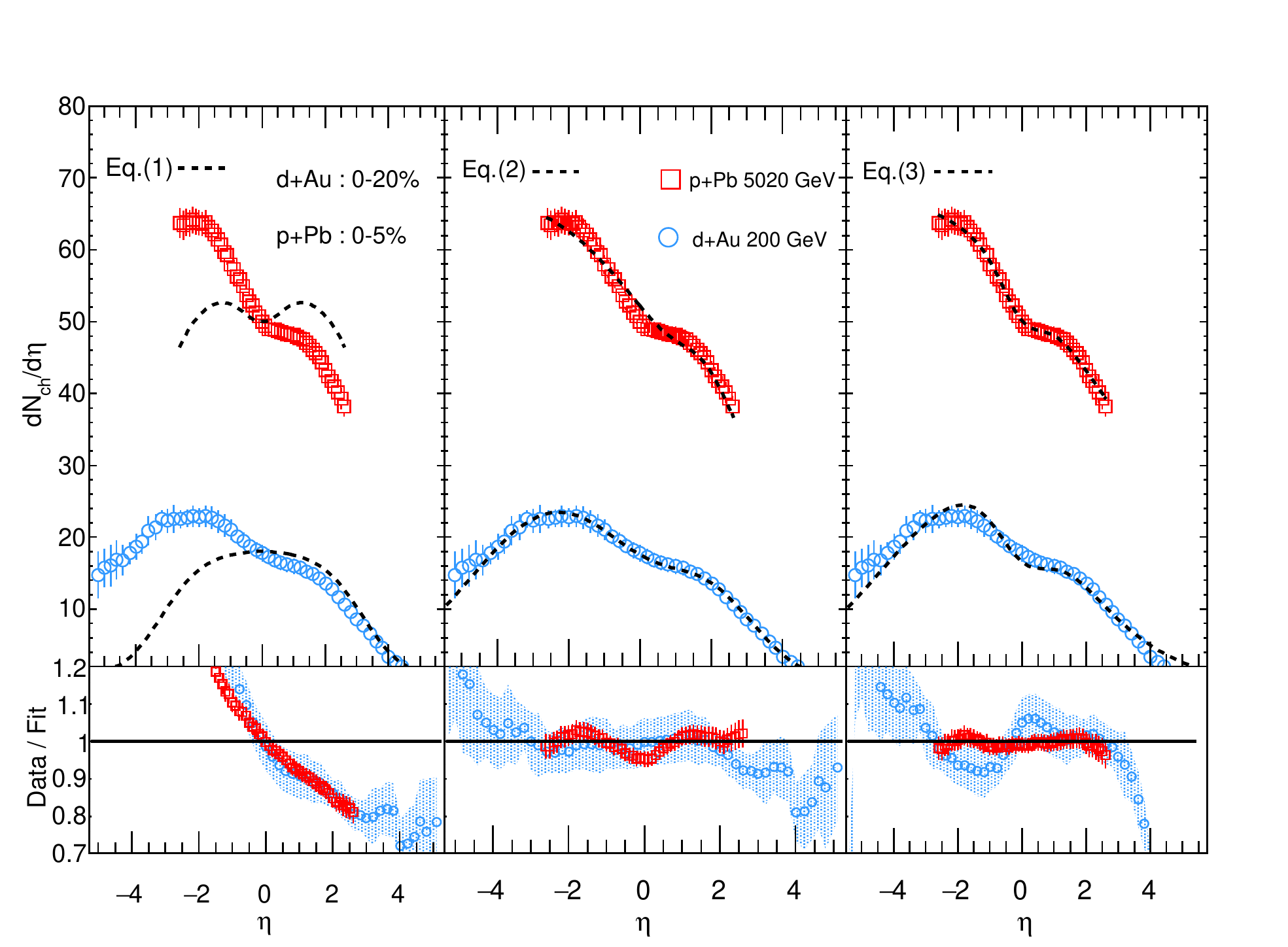}
		\caption{Beam energy dependence of charged 
          particle pseudorapidity density distributions measured in minimum bias  \dAu\ and \pPb\  
          collisions measured by the PHOBOS collaboration~\cite{eta_dAu} and the 
          ATLAS collaboration~\cite{pPb_eta_dist}. Dashed lines show best fits obtained with Eqs.~(\ref{eq:fT}-\ref{eq:fGMG}). 
}
\label{fig:etapA}
\end{figure*}

Equation~(\ref{eq:fT})  is motivated by observed symmetric trapezoidal functions
with a plateau around 
mid-rapidity~\cite{eta_dist_Cu}. It was used by the
PHOBOS collaboration to model their measurements and extract the 
produced total particle multiplicity. 
Although Eq.~(\ref{eq:fT}) adequately reproduces
some of the measured distributions, its built-in symmetry about $\eta=0$
limits its applicability  to symmetric collision systems exclusively. 
Equations~(\ref{eq:fGPG}) and (\ref{eq:fGMG}) involve  sum and differences of two Gaussian distributions, respectively. Equation~(\ref{eq:fT}) features four parameters ($c, \alpha, \beta, a$), while Equation~(\ref{eq:fGPG})
involves  six parameters ($A_1, \mu_1, \sigma_1, A_2, \mu_2$, $\sigma_2$). Equation~(\ref{eq:fGMG}) features five parameters ($A_1, \sigma_1, A_2, \mu, \sigma_2$) but reduces to four for symmetric collisions, which are characterized by $\mu$ = 0. 

Figure~\ref{fig:schematic} shows schematic diagrams of
$dN_{\rm ch}/d\eta$ distributions obtained for pp, \pA, and \AonA\ collisions obtained with  the ans$\ddot{\mathrm a}$tze  embodied by Eqs.~1-3.
  The red and blue lines and associated shaded areas schematically represent the contributions from  nucleon participants originating from  either incoming nuclei.  In the middle panel row, the blue dash line corresponds to the sum of both contributions. The shape of the distribution is here determined by the relative contributions of the incoming nuclei as well as the degree of nuclear stopping achieved in the collisions.  In the bottom panel row, the relative contributions and stopping are modeled with a difference of two Gaussians as per Eq.~(3) and illustrated with the black dash line.  In the upper panel, the trapezoidal ansatz sums  contributions from both incoming nucleai, and is thus  not easy to visualize. In each case, the overlap region can be visualized as a measure of the dip at $\eta=0$ for overall distribution. If the overlap region decreases, then the dip is pronounced and if the overlap region increases, the overall distribution becomes flat.
 
 The three functions are used to fit the available experimental data 
displayed in Figs.~\ref{fig:etapp} -- \ref{fig:etapA}. 
The parameter $\mu$ of  Eq.~(\ref{eq:fGMG}) is set to zero for 
symmetric collisions but left unconstrained for asymmetric systems.  Fits to pseudorapidity distributions measured in pp and \ppbar\ collisions are displayed in Fig.~\ref{fig:etapp}; those to  \CuCu, \AuAu, and \PbPb\ collisions data are shown in Fig.~\ref{fig:etaAA}; whereas those to asymmetric systems, \dAu\ and \pPb\ collisions,
are presented in Fig.~\ref{fig:etapA}. The upper panels of each figure display experimental
data and fits obtained with the three functions in   distinct panels horizontally, while the lower
panels of the figure show ratios of the measured data to the fits. 
The fits  were carried out with  the ROOT least square  
minimization function~\cite{ROOT}. Data uncertainties included in the  
fits were set as quadratic sums of statistical and systematic errors reported by the experiments.   
The goodness  of fit is reported in terms of  $\chi^2$ per 
degrees of freedom ($\chi^2/$NDF) in Tab.~\ref{tab:chi2}. 
\begin{table}[htbp]
	\centering 
	
	\begin{tabular}{@{} ccrccc @{}}  
		\hline
		System  &  Centrality & \sNN  & \multicolumn{2}{c}{$~~~~~\chi^2$/NDF} \\
		&                  & (GeV) &  Eq.~(\ref{eq:fT}) & Eq.~(\ref{eq:fGPG}) &  Eq.~(\ref{eq:fGMG}) \\
		\hline 
		pp      & MB & 900 & 1.056 & 0.552 &  0.826   \\
		pp      & & 2360 & 0.691 & 1.367 & 0.742  \\
		pp      & & 2760 & 2.670 & 14.805 & 1.495  \\
		pp      & & 7000 & 0.458 & 14.805 & 1.495  \\
		pp      & & 8000& 1.103 & 9.320 & 0.157  \\
		pp.     & & 13000& 0.416 &  1.312 & 0.0145    \\
		\hline 
		\ppbar & &630 & 2.355 & 2.636 & 2.144  \\
		\ppbar & &1800 & 0.986 & 0.184 & 0.155 \\
		
		\hline 
		
		CuCu &(0-6\%) & 22.4 & 1.1806   & 1.503  &  1.026  \\
		CuCu & & 62.4 & 0.802 &   0.778 &   0.766  \\
		CuCu & & 200 & 0.877  &  1.095  &  1.185  \\
		\hline 
		AuAu & & 19.6 & 0.596  &  0.725  &  0.592  \\
		AuAu & & 62.4 & 2.184  &  2.074  &  0.412  \\
		AuAu & & 130 & 1.889  &  2.176  &  0.179  \\
		AuAu & & 200 & 1.103  &  0.262  &  0.419  \\
		\hline 
		PbPb &(0-5\%) & 2760 &  1.813  & 1.562  &   0.943 \\
		PbPb & & 5020 &1.319 & 4.216 & 1.462\\
		
		\hline 
		
		dAu & top 5\% &  200  & No Fit &  2.149 & 3.57 \\
		pPb & mult class & 5020 & No Fit &  3.299 & 2.118 \\
		
		\hline 
	\end{tabular}
	\caption{$\chi^2$/NDF of the fits of data presented in 
		Figs.~\ref{fig:etapp}-\ref{fig:etapA} with 
		Eqs.~(\ref{eq:fT}-\ref{eq:fGMG}). MB denotes minimum bias distribution. 
	}
	\label{tab:chi2}
\end{table}

We find  that the three functions fit
the pp data relatively well with $\chi^2/$NDF typically smaller than
 3. However, best fits are achieved with Eqs.~(\ref{eq:fT}) and (\ref{eq:fGMG}).
 Similarly, we find that all three functions 
provide reasonably sensible parameterizations of the \AuAu, \PbPb, and 
 \CuCu\ data presented in Fig.~\ref{fig:etaAA}. We note, however,  that  Eq.~(\ref{eq:fT}) 
 yields  fits with smaller deviations from the data, on average, in the central rapidity region. We also find that Eq.~(\ref{eq:fGPG}) does not 
 fully reproduce the dip structure observed at central rapidity in high-energy datasets. 
   
As anticipated,  fits with Eq.~(\ref{eq:fT}) fail to describe pseudorapidity density distributions measured for asymmetric systems  but  Eqs.~(\ref{eq:fGPG},\ref{eq:fGMG}) produce 
reasonable fits. Note, however, that 
deviations in excess of 5\% are observed at $|\eta|>3.2$ with these models. Irrespective of system size, centrality and collision energy, Eq.~\ref{eq:fGMG} can be used for forward $\eta$ ranges up to the beam rapidity (where experimental measurement has severe limitations) to estimate the number of produced charge particles.  
Altogether, we conclude that Eq.~(\ref{eq:fGMG}) provides the best descriptions of the
pseudorapidity density distributions, regardless of collision system size and energy considered in this work.

\section{Multiplicity distributions at large $\eta$: limiting fragmentation}
\label{sec:limitingFragmentation}

Fits with Eqs.~(1-3) of  pseudorapidity distributions measured in  \PbPb\ collisions 
at 2.76 and 5.02 TeV, discussed in the previous section, successfully model the data but are under constrained at large rapidities. In this section, we use the notion of limiting fragmentation to provide constraints on the shape of these distributions at large rapidity. 
In high-energy hadronic collisions, the limiting 
fragmentation~\cite{benecke,raha,gelis,wolschin1} concept 
stipulates that pseudorapidity densities  reach a 
fixed or universal curve close to the beam rapidity 
($y_{\rm beam}$). 
This implies that the particle production in the 
rest frame of one of the colliding hadrons is (approximately) independent 
of the collision center-of-mass energy. 
Many explanations have 
been suggested to interpret this behavior, including  
gluon saturation in the color glass condensate (CGC) framework~\cite{cgc1, cgc2, cgc3, cgc4}. 
Indeed, parton distribution functions measured in deep inelastic scattering 
show that, at very high collision energies, gluons densities largely dominate those of quarks. 
 This suggests that the medium produced in these collisions 
mostly consists of gluons. With increasing collision energy, the gluon 
density increases, eventually leading to saturation.

In the previous section, we found that Eq.~(\ref{eq:fGMG}) provides
the best fit to the experimental data considered 
in Fig.~\ref{fig:etapp} -- \ref{fig:etapA}. But the
fits remain poorly constrained at large rapidities, i.e.,  at
rapidities in excess of $|\eta|>3.5$. In this context, 
we investigate whether the notion of limiting fragmentation 
can further constrain our modeling of the particle density distributions.  

\begin{figure}[!th]
	\centering\includegraphics[width=0.8\textwidth]{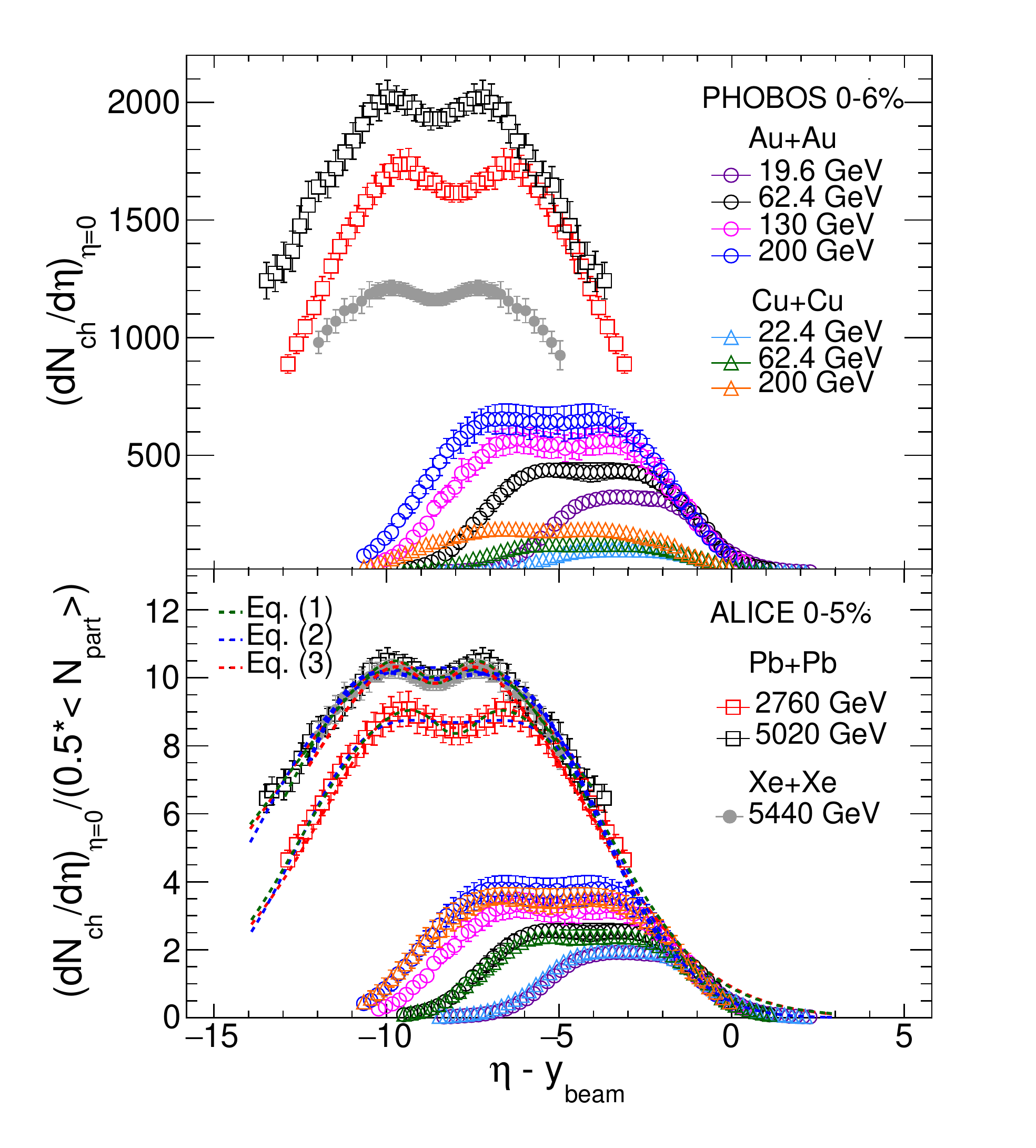}
	\caption{Limiting fragmentation behavior for \AuAu ,
          \CuCu,\ Xe-Xe\ and \PbPb\  collisions at large $\eta$-ranges, plotted
          as a function of $\eta-y_{\rm beam}$. The y-axis in the
          lower panel is scaled by the number of
          participating nucleons pair. }
\label{fig:limiting}
\end{figure} 

Recent studies of limiting fragmentation have shown that Glauber-inspired models of particle production in heavy-ion collisions generally fail to reproduce limiting fragmentation~\cite{Goncalves:2019uod,Sahoo:2018osl} behaviour, especially at LHC energies. These studies indicate that the particle production
is a function of the combination of $N_{\rm part}$ and number of collisions ($N_{\rm coll}$), as the ratio between the two depends non-trivially on the collision energy. Hence, if the nuclei-sized domains are uncorrelated, one generically expects limiting fragmentation to be broken and which is also true in Color Glass type models.
In Ref.~\cite{Goncalves:2019uod}, the authors have argued that wounded parton models, provided the nucleon size and parton density vary predominantly with Bjorken-$x$, could in principle
reproduce both multiplicity dependence with energy and limiting fragmentation. The different calculations can be verified by 
studying the limiting fragmentation behaviour of particle production by re-plotting the 
 the  pseudorapidity density distributions measured in central  \CuCu,
\AuAu, Xe-Xe and \PbPb\ collisions at RHIC and LHC energies as a function of shifted rapidities,  $\eta-y_{\rm beam}$.

The upper panel of Fig.~\ref{fig:limiting} shows the pseudorapidity distributions for central collisions at different colliding energies as a
function of  $\eta-y_{\rm beam}$ for Xe--Xe~\cite{ALICE_XeXe} and Pb--Pb~\cite{ALICE_PbPb} systems at the LHC, and Au--Au collisions at the RHIC energies. 
We observe that the distributions tend to converge
towards a single curve close to $\eta-y_{\rm beam}\sim 0$. This convergence is observed to be dependent on the system size.
This is already quite remarkable considering that the distributions
correspond to systems with rather different number of participants
and  collision energies. 
Accounting for the system sizes, i.e., scaling (dividing) the pseudo-rapidity densities by their respective number of participant pairs, $\la N_{\rm part}\ra/2$, we obtain the distributions shown in  the lower panel of Fig.~\ref{fig:limiting}. 
We observe that the scaled distributions obtained from Xe--Xe, Au--Au, and Pb--Pb collisions at several  energies 
closely overlap  and more or less follow  a  universal limiting fragmentation behavior near $\eta-y_{\rm beam}=0$.

We further test the notion of limiting fragmentation with fits of the data  presented in Fig.~\ref{fig:limiting}  with Eqs.~(\ref{eq:fT}-\ref{eq:fGMG}).  Fits of the different data sets, presented in the figure, indeed merge together near the beam rapidity. In order to further validate the different ans$\ddot{\mathrm a}$tze, the fits were performed by restricting the fit regions and then extrapolating to higher $\eta$. This is verified for Xe-Xe collision (at $\sqrt{s_{\rm NN}} =$ 5.44 TeV) and Pb-Pb collisions (at both $\sqrt{s_{\rm NN}} =$2.76 TeV and $\sqrt{s_{\rm NN}} =$5.02 TeV), by fitting the experimental data in the ranges  (i) $|\eta|\le 2.0$ and   (ii) $-2 < (\eta- y_{\rm beam}) < 3$. We find that the extrapolations of these fits in the beam rapidity are in near perfect agreement, 
with a maximum mutual difference  of  1\%. We also verified that integrals of the fits, yielding total charged-particle multiplicity, differ by less than  5\%. Additionally, we further checked the validity of the 
limiting fragmentation hypothesis by considering fits to two hybrid datasets. These hybrid datesets
were constructed by joining data points from LHC energies    in the range $-13.0 < (\eta-y_{\rm beam}) <  -4.0$ (where  experimental data are available), with $\langle N_{\rm part}\rangle$ scaled values from the STAR 200 GeV data points in the range $-2.0 < (\eta-y_{\rm beam}) < 2.0$. Fits of the two hybrid sets were then performed and we verified that
their integrals matched those of  constrained fits to LHC only data with maximum deviations of  3.5\%.  
We thus conclude that, within the precision afforded with the LHC data, one verifies that  (1) the limiting fragmentation hypothesis is approximately valid at the LHC and  (2) one can then exploit the hypothesis to constrain the LHC data at large rapidity. Using this limiting fragmentation hypothesis, and extrapolating  fitting functions to beam rapidity, it is thus possible to estimate, with reasonable accuracy,   the total charge particles ($N_{\rm ch}^{\rm total}$) production
at LHC energies and compare with values obtained at RHIC energies. We discuss the  extraction of $N_{\rm ch}^{\rm total}$ in detail in the next section.

\section{Total charged-particle multiplicity}
\label{sec:TotalMult}

\begin{figure}[!th]
\centering\includegraphics[width=0.8\textwidth]{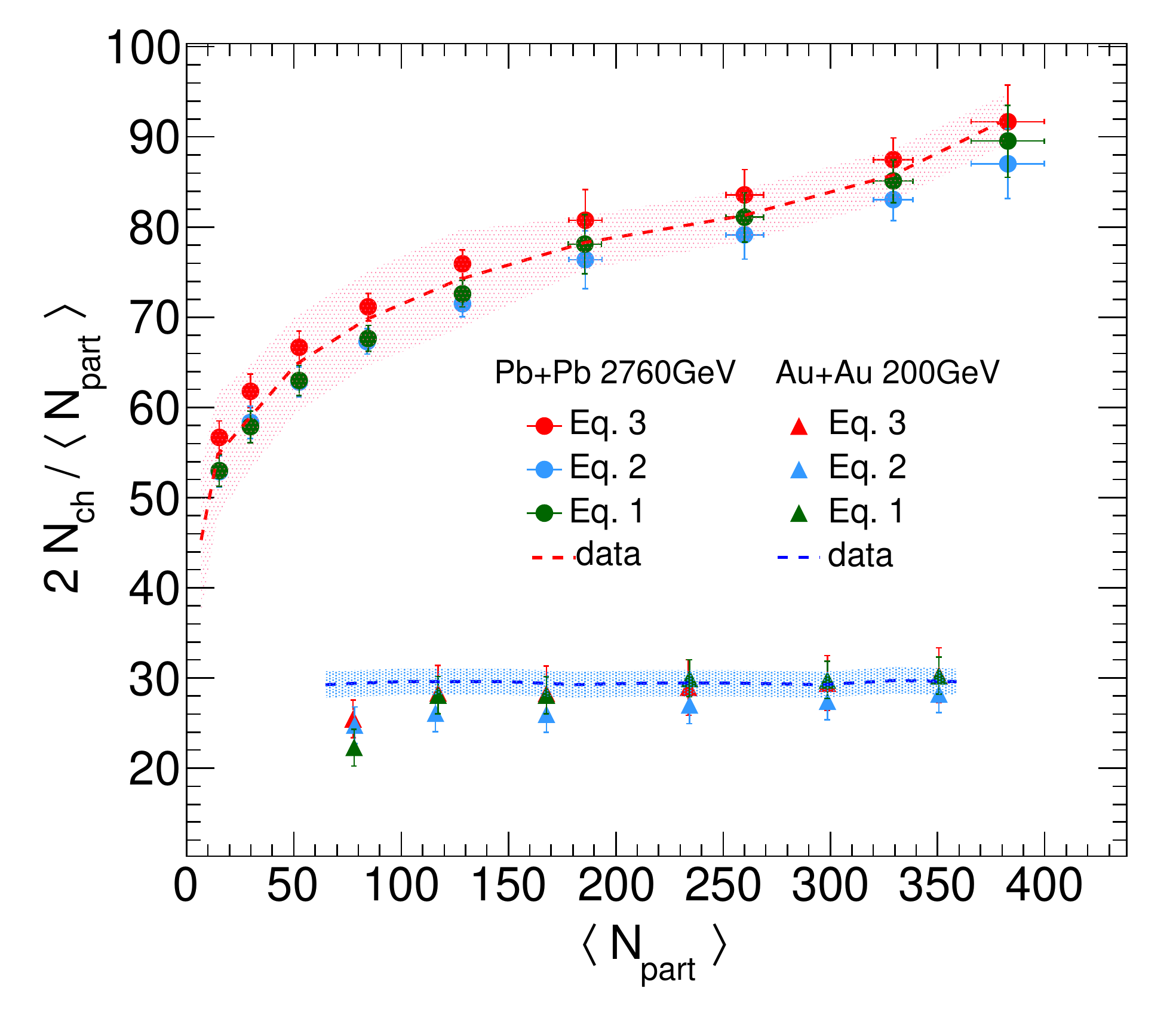}
\caption{
Total charge particle multiplicity scaled  by the number of participant pair, $\la N_{\rm part}\ra / 2$, as a 
function of $\la N_{\rm part}\ra$ based on Eqs.~(\ref{eq:fT}-\ref{eq:fGMG}). 
Red and blue dash lines correspond to data reported by the ALICE and PHOBOS collaboration based on measured charge particle multiplicity measured in the range $|\eta| \le 5.5$  and extrapolated to
 $-y_{\rm beam}~\le~\eta~\le+y_{\rm beam}$. The shaded bands represent error bars corresponding to the correlated systematic uncertainties reported by the experiments~\cite{alice-nch,phobos-nch}. }
\label{Nch_check}
\end{figure}

We  proceed to 
determine the total  charged-particle  multiplicity, $N_{\rm ch}$,
produced in  \CuCu, \AuAu, and \PbPb\ collisions by integration of the
fitted pseudorapidity densities, constrained by limiting
fragmentation,  over the full range of particle production.  
Figure~\ref{Nch_check}  presents  values of $N_{\rm ch}$
scaled by $\la N_{\rm part}\ra / 2$ as a function of $\la N_{\rm part}\ra$ for  \PbPb\ collisions
at 2.76 TeV and \AuAu\ collisions at 200 GeV. Experimental data points reported
by the ALICE\cite{alice-nch} and PHOBOS\cite{phobos-nch} collaborations are
shown with red and blue dash curves, respectively. Total charged-particle production values are obtained by integration of the fitted Eq. (1-3) in the range $-y_{\rm beam} \le \eta \le y_{\rm beam}$. Values obtained with Eqs. (1), (2), and (3)  are shown with solid red, blue and green points, respectively.

We find that the scaled values of $N_{\rm ch}$ (red triangles and red circles)  obtained by integration of Eq.~(\ref{eq:fGMG}) are consistent, within  uncertainties (represented by shaded bands), with those reported by the PHOBOS and ALICE collaborations. Only
 the $N_{\rm ch}$ values obtained at the  lowest $\la N_{\rm part}\ra$ shown appreciably underestimate the PHOBOS data. 
Scaled values of $N_{\rm ch}$ obtained by integration of  Eq.~(1) follow a similar trend while 
those obtained with Eq.~(2) tend to systematically underestimate the values reported by PHOBOS.  Overall, we find the best agreement with PHOBOS data  is achieved using  Eq. (3),  with deviations of the order of 0.5\% compared to 1\%  with the other two equations. 
Hereafter, we use the differences of the three fit extractions as an estimate of the systematic errors associated with the extrapolation procedure based 
 on fits of Eq.~(3) to obtain the total charged-particle multiplicities. 

\begin{figure}[!th]
	\centering\includegraphics[width=0.8\textwidth]{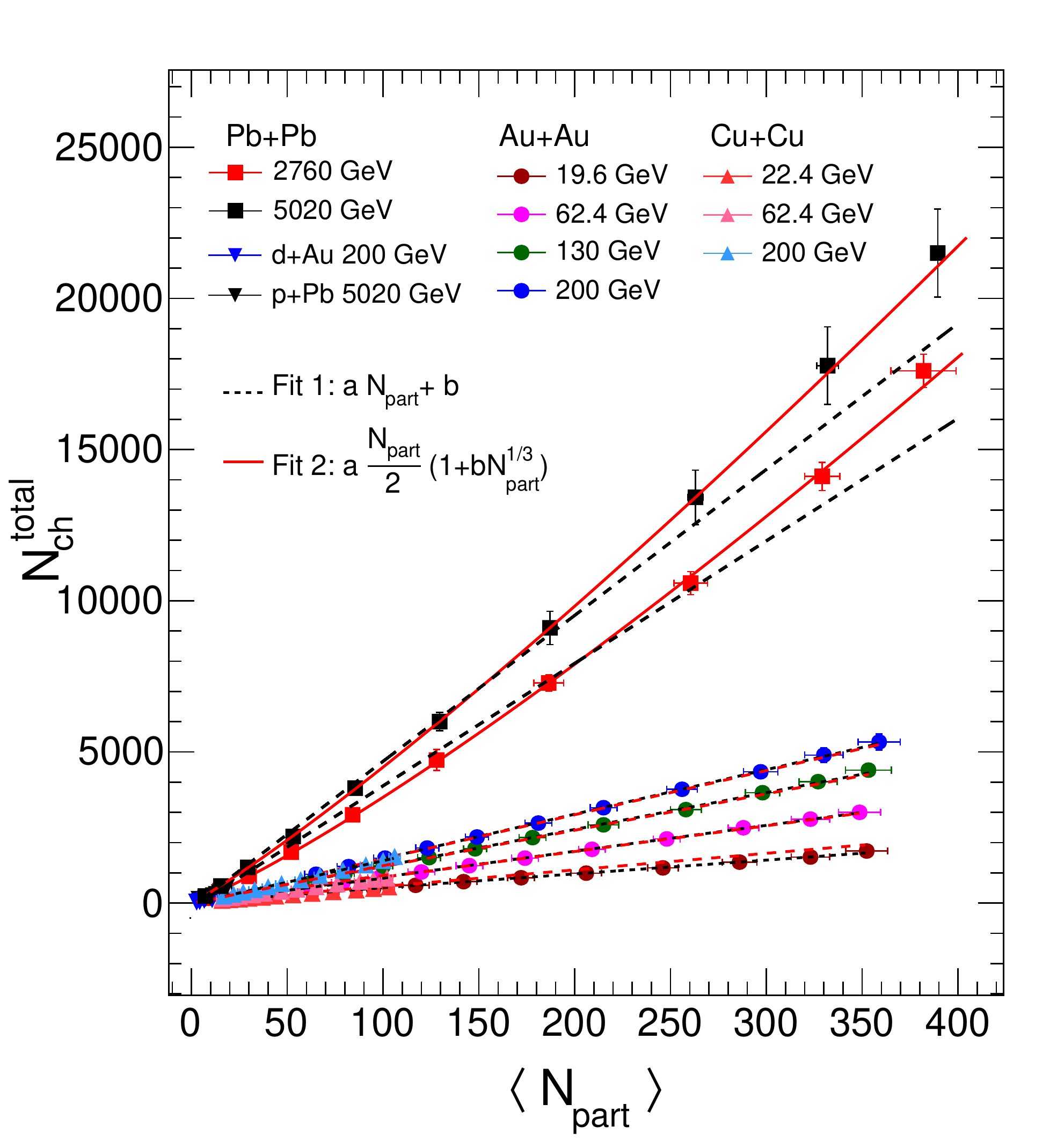}
	\caption{Centrality dependence of the total charged-particle 
          multiplicity, estimated from integrals of Eq.~(\ref{eq:fGMG}) across the  range $-y_{\rm beam}~\le~\eta~\le+y_{\rm beam}$, in pp, \dAu, \pPb, \CuCu, \AuAu,  and \PbPb\ 
          collisions at RHIC and LHC energies.
          }
    \label{Nchtotal}
\end{figure}

We next proceed to use fits of the measured pseudorapidity distributions with Eq.~(\ref{eq:fGMG})  to extract
values of $N_{\rm ch}$ for several colliding systems, collision
energies, and collision centralities. Results  are shown in Fig.~\ref{Nchtotal}  as a function of the number of participating
nucleons in pp collisions at 19.6~GeV, 200~GeV and 2.76~TeV, \AuAu\  collisions at 19.6, 62.4, 130, and 200 GeV, \CuCu\ collisions at 22.4, 62.4 and 200 GeV, \dAu\ collisions at 200 GeV, \PbPb\ collisions at 2.76 and
5.02 TeV, and \pPb\ collisions at 5.02 TeV.  We observe that the integrated multiplicities 
generally exhibit a power law dependence on the average number of participants. Additionally, while the
integrated multiplicities obviously increase with the system size and collision energy, 
they otherwise appear, upon first inspection,  to feature a similar  power-law 
dependence on $N_{\rm part}$. 

We further examine the $N_{\rm ch}$ dependence on $N_{\rm part}$ 
by considering  parameterizations of this dependence with (a) a linear function  $a N_{\rm
   part}+b$, and (b) a power law $a \frac{N_{\rm part}}{2}  (1+bN_{\rm part}^{\frac{1}{3}})$, 
 shown in Fig.~\ref{Nchtotal} with black dashed and red solid lines, respectively. 
 We find that the power-law parameterization provides a better  description of the evolution of  $N_{\rm ch}^{\rm total}$ with $N_{\rm part}$.
Notably, the linear fit fails to describe the evolution of $N_{\rm
  ch}^{\rm total}$ with $N_{\rm part}$ at LHC energies. Deviations are observed for
peripheral collisions with both parameterizations. Moreover,  
both the linear and power law functions provide 
a rather poor description of the computed multiplicities in the case of \pPb~collisions. 

\begin{figure}[!th]
		\centering \includegraphics[width=0.8\textwidth]{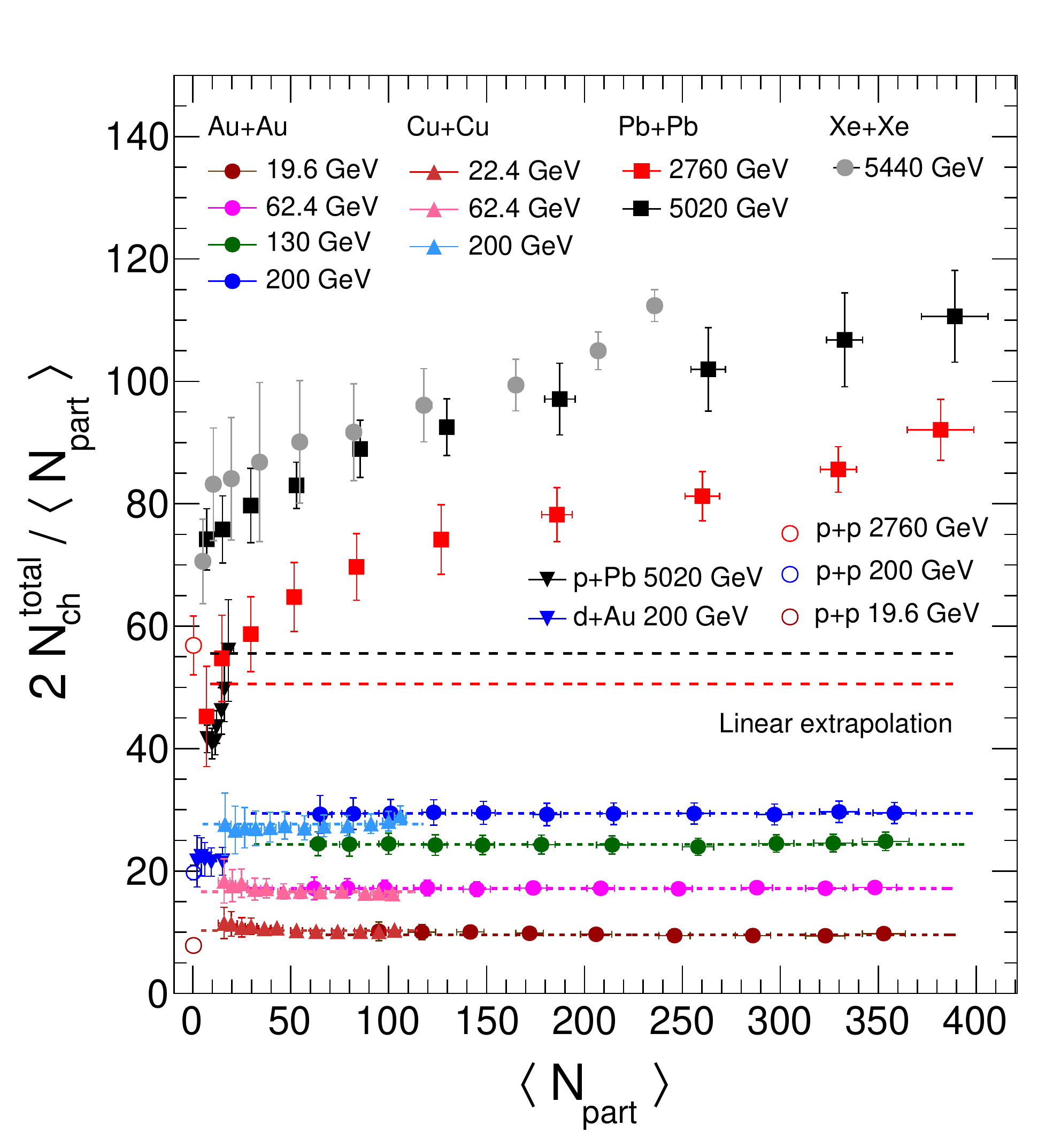}
	\caption{Centrality dependence of total charged multiplicity 
          per participant pair in pp, \dAu, \pPb, \CuCu, \AuAu,  and \PbPb\  
          collisions at RHIC and LHC energies~\cite{Phobos_data_all, eta_dist_2760, Pb_5020_eta}. 
}
    \label{NchtotalVsNpart}
\end{figure}

\begin{figure}[!th]
		\centering \includegraphics[width=0.75\textwidth]{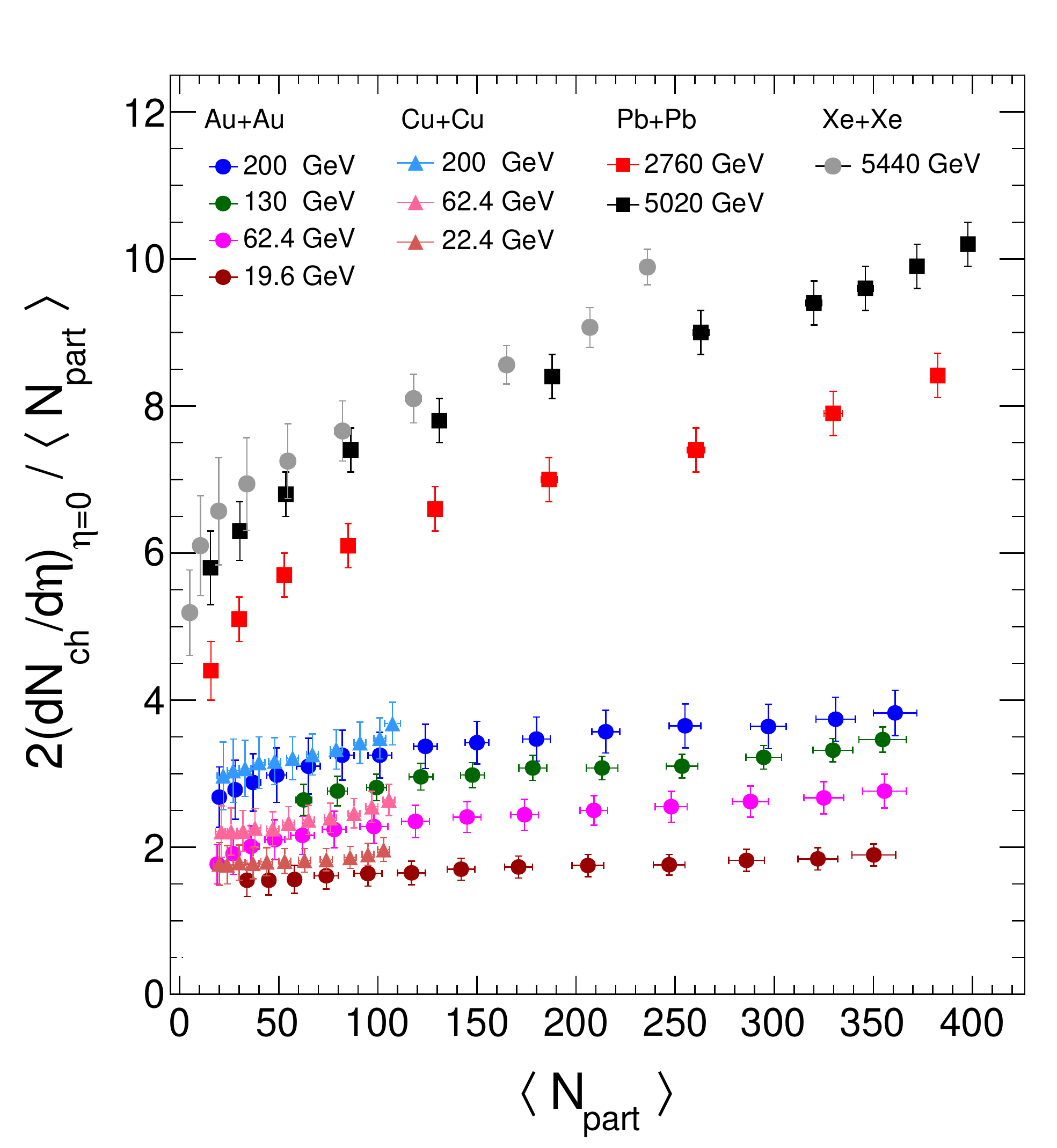}
		\caption{Centrality dependence of charged-particle multiplicity 
          density at mid-rapidity in  \CuCu, \AuAu, \PbPb\ and Xe-Xe  collisions 
            at RHIC and LHC energies. 
          }
\label{NchVsNpart}
\end{figure}

In order to further examine the evolution  of $N_{\rm ch}^{\rm  total}$ with $N_{\rm part}$, 
we plot the centrality dependence of total charged
particle multiplicities scaled by the number of  participant pair
in Fig.~\ref{NchtotalVsNpart}.   We observe that for \CuCu\ and \AuAu\ collisions at RHIC
energies, scaled  values of 
$N_{\rm ch}^{\rm total}/(\langle N_{\rm part}\rangle/2)$ are  essentially independent of the collision centrality,
whereas $(dN_{\rm ch}/d\eta)_{\eta=0}/(\langle N_{\rm part}\rangle/2)$, plotted in Fig.~\ref{NchVsNpart}, displays a
monotonic rise with $N_{\rm  part}$ in these collision systems. 
This implies that the shape of the $\eta$ density distribution changes with
centrality and becomes more peaked with increasing centrality.
In contrast, we find that, at  LHC energies,
both $N_{\rm ch}^{\rm total}/(\langle N_{\rm part}\rangle/2)$ (Fig.~\ref{NchtotalVsNpart}) and $(dN_{\rm
  ch}/d\eta)_{\eta=0}/(\langle N_{\rm part}\rangle/2)$ (Fig.~\ref{NchVsNpart}) display monotonic increase with 
  $N_{\rm part}$. 
For LHC collisions, the ratio $N_{\rm ch}/(\langle N_{\rm part}\rangle/2)$ shows a growth,
compatible with a power-law behavior. A similar behavior is 
observed for \pPb\ collisions at 5.02~TeV (Fig.~\ref{NchtotalVsNpart}).

The observed  violation of participant scaling at LHC energies  is in sharp contrast to the near perfect scaling 
observed at  RHIC energies. Furthermore, a scaling violation is
observed for both charged-particle multiplicity density at
mid-rapidity as well as the total number of charged particles. The
causes of these violations can be manifolded. First, the increase in
beam energy by more than one order of magnitude from RHIC to  LHC
energies makes the typical Bjorken-$x$ at LHC much lower compared to
that at RHIC. At RHIC energies, a transverse mass, $m_{\rm T}$, of 1
GeV corresponds to $x\sim 10^{-2}$ at $y=0$, whereas at LHC it
corresponds to $x\sim 10^{-4}$. Bjorken-$x$ values are even lower at
large $\eta$. The gluon density is expected to grow and reach saturation
with lowering $x$~\cite{parton_dist}.   At the LHC, one gets to the
small $x$ domain where gluon productions dominates thereby  producing
large number of additional particles with no relation to the number of
participants. This is consistent with the CGC formalism of the initial
state of the colliding nuclei.

Alternatively,  particle production  at high energy
may be described in terms of a two component model involving soft and hard components,
$\sigma_{\rm total} = \sigma_{\rm soft} + \sigma_{\rm hard}$, in which 
$\sigma_{\rm soft}$ represents the cross-section for soft particle production
and is proportional to $N_{\rm part}$, whereas $\sigma_{\rm hard}$, the cross-section for high-$p_{\rm T}$ particle production,  is
proportional to the number of inelastic nucleon-nucleon collisions ($N_{\rm coll}$). A significant 
increase of $\sigma_{\rm hard}$ from RHIC to LHC, relative to $\sigma_{\rm soft}$ could 
then possibly explain the observed departure from  $N_{\rm  part}$ scaling.

\subsection{Extrapolation of particle multiplicities to lower beam energies}

\begin{figure}[!th]
\centering \includegraphics[width=0.8\textwidth]{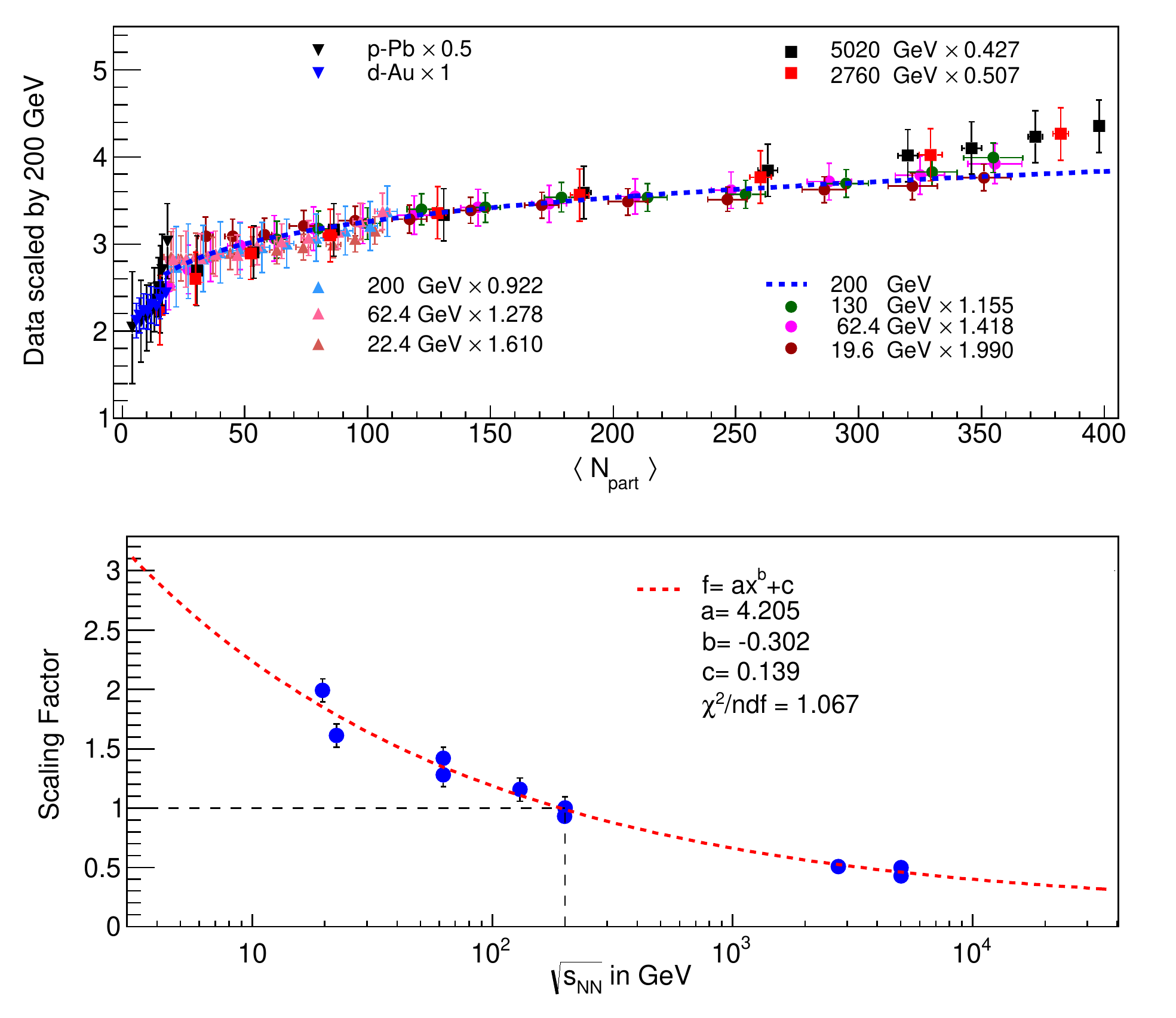}
\caption{ (Upper) Centrality dependence of charged-particle 
          multiplicity density scaled to that of AuAu collisions at
          \sNN~=~200 GeV. 
(Lower)  Scaling factors for charged-particle multiplicity density to
the data at 200~GeV. 
}
\label{scale200}
\end{figure}

We  use  the power law obtained in the previous section to ``predict"
the total charged-particle production as a function of the number of participants at the FAIR and NICA facilities,
expected to become online in 2022 and 2025, respectively. To calculate these predictions, we first remark that the shape of the $\langle N_{\rm part}\rangle$ dependence of the central rapidity particle density for RHIC energies is essentially invariant with respect to \sNN. To illustrate this approximate invariance, we plot central multiplicity densities scaled to the corresponding multiplicity density at \sNN~=~200 GeV as a function of $\langle N_{\rm part}\rangle$ for several collision systems and energies in Fig.~\ref{scale200}. 
The scaling factors were determined as the ratio of multiplicity density at central rapidity measured at   different beam energies  \sNN\ to the multiplicity density observed at central rapidity in \sNN~=~200 GeV \AuAu\ collisions. These are listed 
for each collision system and energy in the upper panel of the figure. The scaled densities are compared to the CGC initial condition motivated fit (discussed in the next section) to the data at \sNN~=~200 GeV, shown as a blue dash line.

\begin{figure}[!th]
		\centering \includegraphics[width=0.75\textwidth]{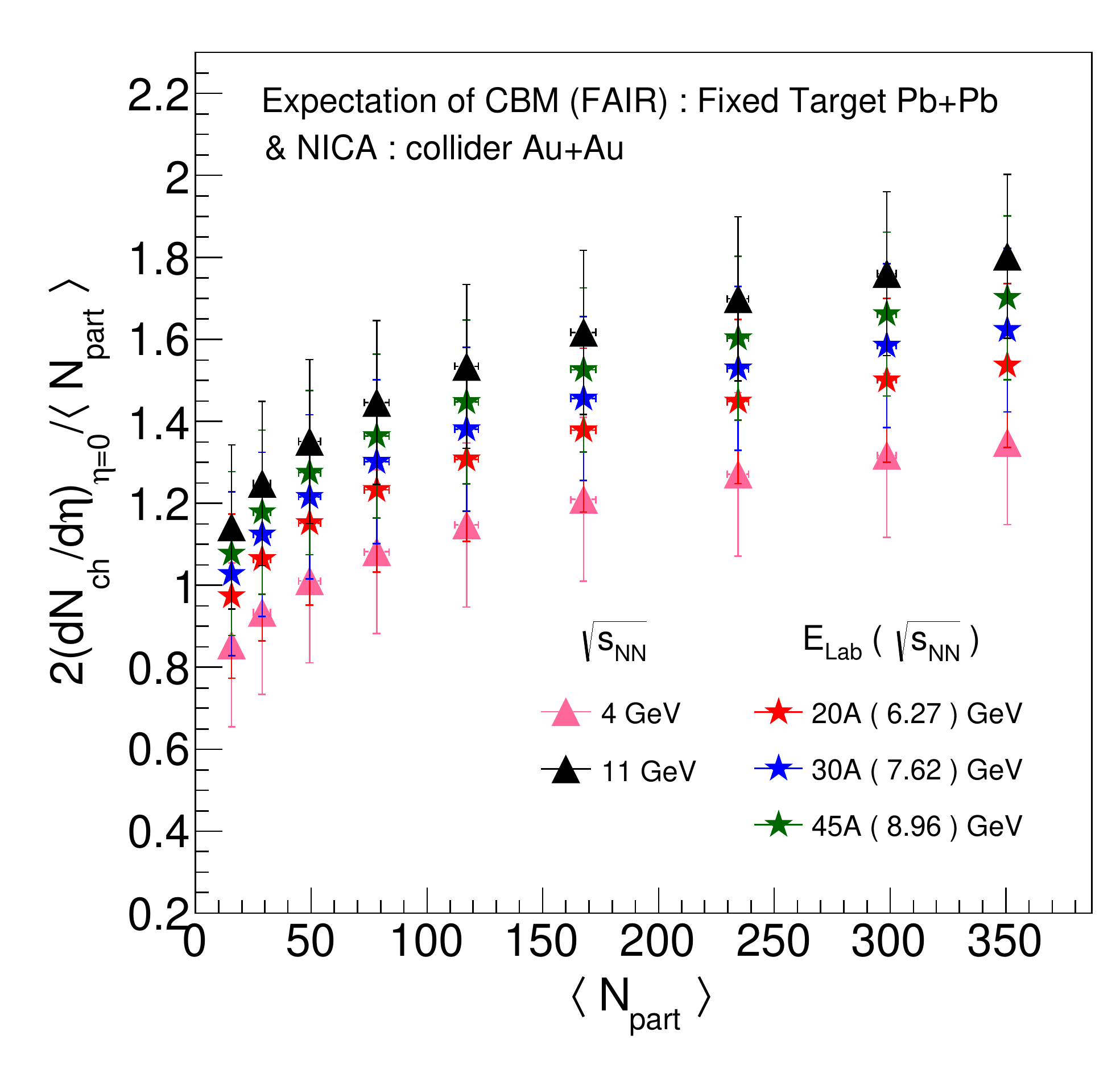}
	\caption{Expected evolution of charged-particle multiplicity density with centrality for CBM (FAIR) energies.}
\label{lowenergy2}
\end{figure}

We observe from Fig.~\ref{scale200} that the overlap of the data points is reasonable at energies lower than \sNN~=~200 GeV, which makes it possible to predict the particle densities at lower collision energies.
The scaling factors are plotted as a function of \sNN~ in the lower panel of the figure and fitted with a first order  polynomial shown by the red dash line. We extract the coefficients $a$ and $b$, and use these 
to obtain scaling factors for NICA and FAIR energies.
These scaling factors are used to obtain predictions of 
collision centrality evolution of the central particle density per participant, 
$\left. dN_{\rm ch}/d\eta\right|_{\eta=0}/\langle N_{\rm part}\rangle/2$. This is shown in Fig.~\ref{lowenergy2} as a function of $\langle N_{\rm part} \rangle$.


\subsection{Extrapolation of particle multiplicities to higher beam energies}

The High-Energy Large Hadron Collider (HE-LHC)~\cite{he-lhc} and the  FCC~\cite{fcc} accelerators proposed at CERN will achieve unprecedented large collision energies for pp as well as heavy-ions. The expected energies for Pb-Pb collisions are 11~TeV and 39~TeV for HE-LHC and FCC, respectively. 
 It is thus imperative to make predictions for the number of produced particles at such large energies. The scaling technique used to extrapolate the particle multiplicities for collision energies lower than \sNN~=~200~GeV is not appropriate for extending to higher energies as the approximate $N_{\rm part}$ scaling is broken (as per figure~\ref{NchVsNpart}). The indication of the scale breaking for multiplicity density at mid-rapidity is also evident by a closer look to upper panel of Fig.~\ref{scale200} at \sNN~=~2760~GeV and\sNN~=~5020~GeV for $N_{\rm part} > 300$. 
 But using the power-law dependence of  
 beam energy (Fig.~1) for AA collision at top 5\% centrality 
 ($= 0.77~s_{NN}^{0.153\pm0.002}$), we can predict the charged particle multiplicity densities at mid-rapidity for Pb--Pb collisions at 11~TeV and 39~TeV. The extrapolation gives 
 $\frac{2}{\langle N_{\rm part}\rangle}\left. dN_{\rm ch}/d\eta \right|_{\eta=0}$ as $13.279\pm0.504$ and $19.559\pm0.845$, respectively for top central collisions. Taking these values into account, the charged particle multiplicity density at $\eta=0$ are $~\approx2456\pm93$ and $~\approx 3618\pm156$ for 11~TeV and 39~TeV, respectively. As these higher energies probe more low-x region, one should be careful by considering the present knowledge of gluon saturation picture, which could limit the particle production in these energies and push the multiplicity towards a lower value than expected from these extrapolation.

\section{Multiplicity density from initial condition motivated models  }
\label{sec:InitialConditions}

\begin{figure*}[!th]
	\centering \includegraphics[width=1.0\textwidth]{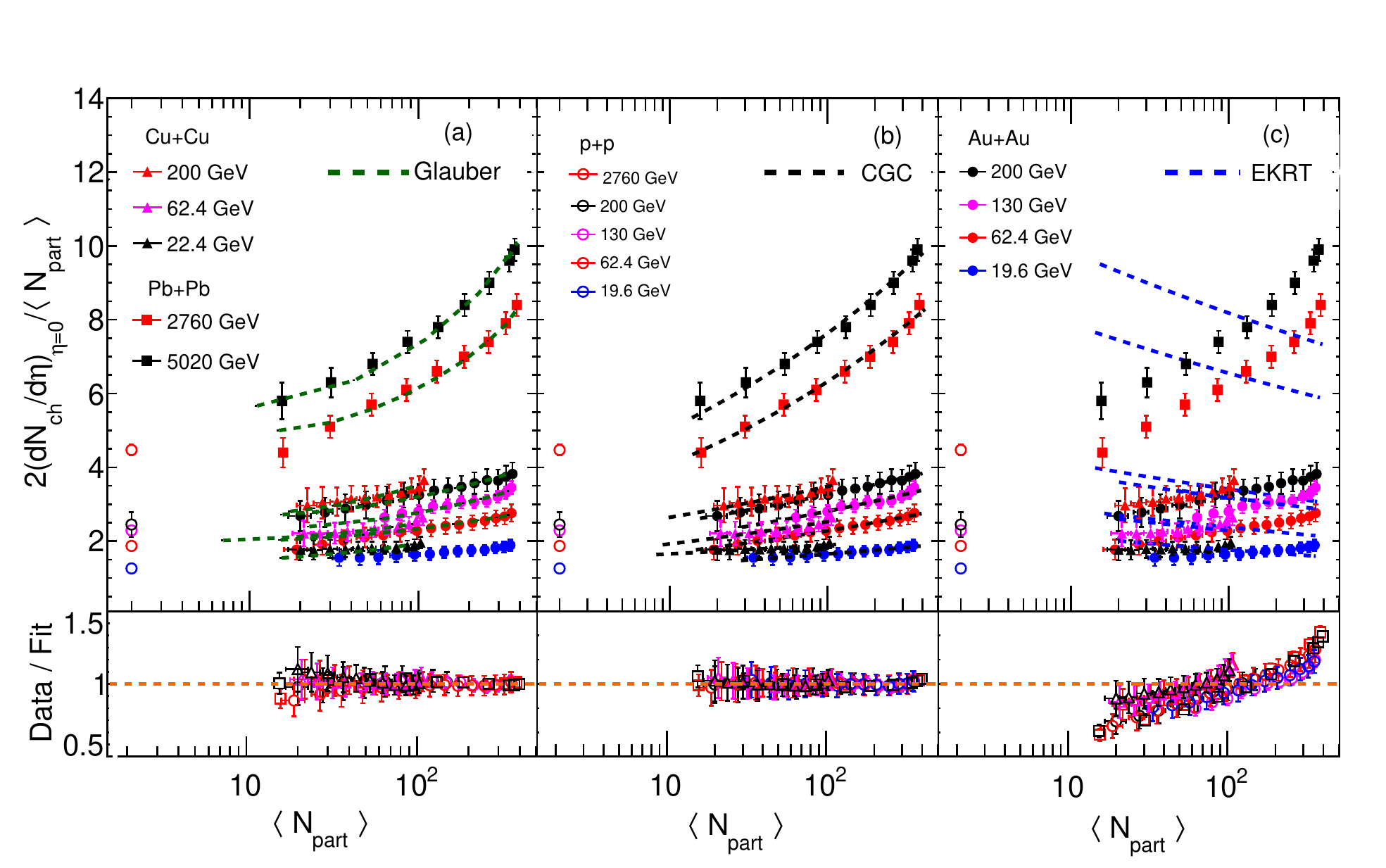}
	\caption{Parameterization  of the $N_{\rm part}$ dependence of charged-particle multiplicity density per participant pair for symmetric collision systems fitted with initial conditions according to  (a) Glauber, (b) CGC, and  (c)  EKRT models.}
\label{initial_cgc}	
\end{figure*}

The collision centrality dependence of the ratio  $N_{\rm ch}/\frac{\langle N_{\rm part}\rangle}{2}$ is expected to be somewhat sensitive to the  initial state 
conditions of  heavy-ion collisions~\cite{alba1,alba2}. 
The measured evolution of charged-particle multiplicity distributions vs. collision centrality,  presented in Fig.~\ref{initial_cgc} for selected 
collision systems, may thus be used to 
contrast predictions obtained with different models. We focus our discussion on the Glauber~\cite{systematic_star,systematic_phobos} and  color glass condensate~\cite{alba1,alba2} models. 

Within the Glauber model, a  soft/hard two-component model is used to parameterize the particle 
production as a function of collision centrality according to 
\begin{eqnarray}\label{eq:twoComponents}
\frac{dN_{ch}}{d\eta}\Big|_{AA}=n_{pp} \left[\left(1-x \right) \frac{N_{\rm part}}{2} +xN_{\rm coll}\right],
\end{eqnarray}
where $N_{\rm part}$ and $N_{\rm coll}$ represent the number of soft and hard 
scatterings, respectively,   and $n_{pp}$ denotes the average number
of produced charged particles per unit pseudorapidity in pp collisions. The variable $x$, representing the fraction of hard collisions, is here 
considered a fit parameter. The fit results of hard scattering component $x$, is within the range of 0.10 to 0.16 and in agreement with previous measurements. Panel (a)  of Fig.~\ref{initial_cgc} displays fits (green dash lines) of  data 
from \CuCu, \AuAu, and \PbPb\ collisions across a wide span of beam energies. To carry out the fits, we evaluated  values of $n_{pp}$ vs. $\sqrt{s}$ based on the parameterization, $n_{pp}$~$\propto~s_{NN}^{0.11}$, presented  for (NSD) \pp\ collisions in Fig.~\ref{fig:model1}.

\begin{figure}[!th]
\centering \includegraphics[width=0.75\textwidth]{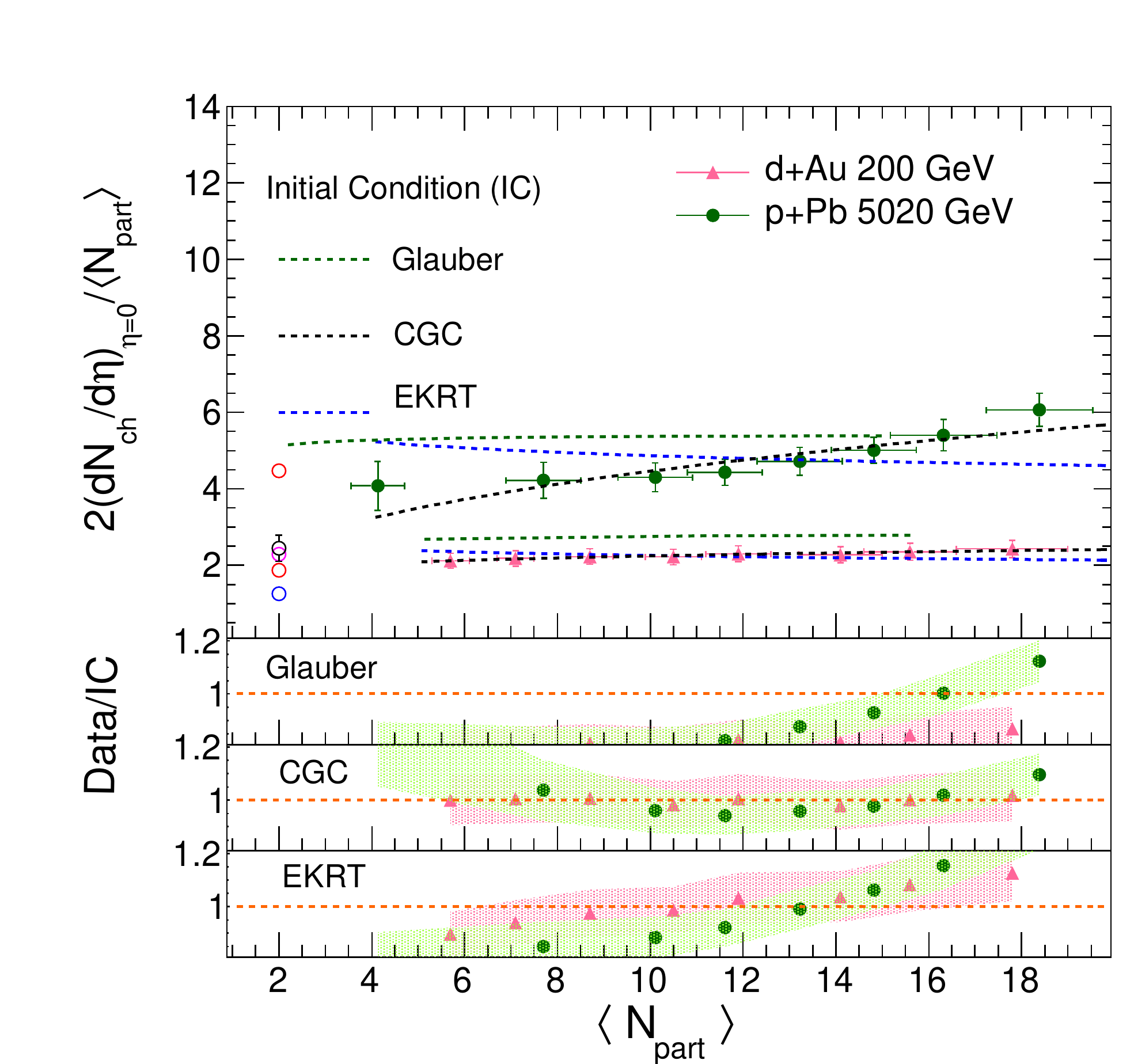}
\caption{d+Au and p+Pb  asymmetric collisions fitted with different 
  initial conditions according to Glauber, CGC and EKRT models.}
\label{initial2}	
\end{figure}

In the context of the Color Glass Condensate model, one expects that small $x$ gluons overlap and 
recombine  thereby reducing the overall number of gluons and   the number 
of hadrons they hadronize into.  The charged-particle density is hence modeled according 
 to~\cite{cgc2}:
\begin{eqnarray}
\label{eq:cgcInit}
\frac{dN_{ch}}{d\eta}\approx N_{part}^{\alpha} (\sqrt{s_{NN}})^\gamma,
\end{eqnarray}
where $\alpha$ and $\gamma$ are free parameters. Fits based on this model are shown in 
Fig.~\ref{initial_cgc} (b). By contrast, models based on final state gluon saturation,
e.g., EKRT~\cite{EKRT2000}, 
predict a decreasing trend in charged-particle multiplicities per participant
nucleon with increasing collision centrality according to 
\begin{eqnarray}\label{eq:ekrt}
\frac{dN_{ch}}{d\eta}=C\frac{2}{3}1.16 \Big(\frac{N_{part}}{2}\Big)^{0.92}(\sqrt{s_{NN}})^{0.4},
\end{eqnarray}
where $C$ is the only free parameter. While the Glauber and CGC initial conditions parameterizations    
shown in panels (a) and (b) provide excellent agreement with measured data, one finds fits based on Eq.~(\ref{eq:ekrt}), presented in Fig.~\ref{initial_cgc} (c)  are in stark disagreement with the data, owing evidently to the fixed $N_{part}$ power smaller than unity. 

We extend this study to \dAu\ and \pPb\ collision systems  in Fig.~\ref{initial2} using the parameterizations~(\ref{eq:twoComponents}-\ref{eq:ekrt}). We find that, in these two systems, the soft/hard two-component model and the EKRT Eq.~(\ref{eq:ekrt}) provide a relatively poor representation of the data. Overall then, we conclude the CGC inspired parameterization, Eq.~(\ref{eq:cgcInit}), provides 
a suitable description of the evolution of the charged-particle
multiplicity density with $N_{\rm part}$ in both symmetric and asymmetric collision systems.

However, we note that recent event-by-event calculations carried out 
using next-to-leading order
EKRT model~\cite{Eskola:2015uda}, with  saturation for soft particle production and viscous hydrodynamics for
the space-time evolution of the produced matter, can well describe the multiplicity  density discussed above.
In addition, the recent theoretical development on initial conditions known as TRENTO~\cite{bass_trento, Shen:2014vra} initial conditions also provides a successful description of the densities (as well as several other observables)  in  \pp, \pPb, \AuAu,  and  \PbPb\  collisions   both at  RHIC and  LHC energies.

\section{Summary}
\label{sec:summary}

We have presented a comprehensive study of the multiplicity and pseudorapidity distributions of
charged particles produced  in \pp, \pPb, \dAu, \CuCu, \AuAu,  and  \PbPb\ collisions 
at energies ranging from a few GeV to several TeV, corresponding to
the available experimental data at RHIC and LHC. The experimental data
have been compared to calculations of selected event generators, including   
PYTHIA, EPOS, AMPT, UrQMD, and THERMINATOR, which feature 
different physics model assumptions.
We find these event generators
qualitatively reproduce the observed particle densities at $|\eta| = 0$. However, none
 are able to satisfactorily explain measured
distributions over a broad range of pseudorapidities. 
With the goal of extrapolating the 
measured data
to forward rapidities, to estimate the total charged particle
production in various collision systems, and to obtain the dependence
on the collision energy, we have studied three different functional
forms to describe the experimental data on the pseudorapidity distributions.
Among these functional forms, the difference of two Gaussian
distributions,  
Eq.~(\ref{eq:fGMG}), is found to best reproduce the measured multiplicity
densities observed in  different collision systems and collision energies.

Furthermore, we  used  Eq.~(\ref{eq:fGMG}) to estimate
the total charged-particle production and 
study the evolution of multiplicity density at central rapidity ($\left. dN_{\rm ch}/d\eta/\langle N_{\rm
    part}\rangle/2\right|_{\eta=0}$) as a function of collision
centrality and collision energy. 
At beam energies $\sqrt{s_{\rm NN}} \le 200$~GeV, the charged-particle rapidity density 
exhibits a modest increase with $\langle N_{\rm part}\rangle$ while the total charge production 
is approximately independent of collision centrality.  
In contrast, at LHC energies, both the particle density at
mid-rapidity and the total charge particle production exhibit a rapid increase with $\langle N_{\rm part}\rangle$. We thus 
conclude that there is a qualitative change in the particle production
at LHC relative to RHIC. At  RHIC energies, 
the multiplicity density at mid-rapidity increases with  $\langle
N_{\rm part}\rangle$ while the total particle production remains
fixed. That implies the  pseudorapidity distribution narrows with
increasing $\langle N_{\rm part}\rangle$
thereby yielding a larger central rapidity density albeit with a fixed
integral. At the LHC, by contrast, both the central rapidity density
and the total charged-particle production increase with 
$\langle N_{\rm part}\rangle$. One then has entered a different regime
of particle production in which both the central rapidity and total
multiplicities per participant monotonically increase with $\langle
N_{\rm part}\rangle$. 

We found that the limiting fragmentation hypothesis holds at 
the TeV energy scale and thus can be used to approximately constrain 
the shape of $dN/d\eta$ distributions and their integrals over the 
full range of particle production. 
In addition, we have studied charged-particle multiplicity 
productions considering different initial conditions.  
We observe that CGC like initial condition is best suited to describe
the published data for both symmetric and asymmetric type of collisions.
We have extended the particle production studies to lower collision
energies corresponding to those of upcoming accelerator facilities of
FAIR at GSI, Darmstadt and NICA at JINR Dubna.
We have extrapolated the charged 
particle multiplicity densities at $\eta=0$
for expected heavy-ion collisions at the proposed 
HE-LHC and  FCC at CERN.

\bigskip
\noindent
{\bf Acknowledgments:} \\
SB \& CP  acknowledge the financial support by the U.S. Department of Energy Office of Science, Office of Nuclear Physics
under  Award  Number  DE-FG02-92ER-40713.


\bigskip
\bigskip
\noindent

{\bf References}
\medskip

\begin{thebibliography}{}

	\bibitem{bialas} A. Biallas, M. Bleszynski, and W. Czyz, Nucl. Phys. B
	  {\bf 111}, 461 (1976). 
	
	\bibitem{bjorken1}  J. D. Bjorken, Phys. Rev. D \textbf{27}, 140 
	(1983). 
	
	\bibitem{kharzeev} D. Kharzeev and M. Nardi,  Phys. Lett. B
	\textbf{507}, 121 (2001). 
	
	\bibitem{jan} J.F. Grosse-Oetringhaus and K. Reygers, J. Phys. G:
	  Nucl. Part. Phys. \textbf{37}, 083001 (2010).


	\bibitem{Phobos_data_all}
	B.~Alver {\it et al.} [PHOBOS Collaboration],Phys.\ Rev.\ {\bf C 83} (2011) 024913. 
	  
	\bibitem{star_bulk} L. Adamczyk {\it et al.} (STAR Collaboration) 
	Phys. Rev. C {\bf 96}, 044904 (2017).
	
	\bibitem{alice_bulk} A. Toia {\it et al.} (ALICE Collaboration)
	J. Phys. G {\bf 38}, 124007 (2011).
	
	\bibitem{raghu1} R. Sahoo,  A. N. Mishra,  N. K. Behera and B. K. Nandi, 
	Adv. high-energy Phys. \textbf{2015}, 612390 (2015).
	
	\bibitem{raghu2}  E. K. G. Sarkisyan, A. N. Mishra, R. Sahoo and A. S. Sakharov, 
	Phys. Rev. D \textbf{93}, 054046 (2016).

	\bibitem{Sumit} 
	S. Basu, T.K. Nayak, K. Datta, Phys. Rev. C \textbf{93}, 064902 (2016). 
		
		\bibitem{pythia1} 
	Torbjorn Sj{\"o}strand, Stephen Mrenna, and Peter Skands, J. High Energy Phys. {\bf05}, 026 (2006).
	
	\bibitem{herwig} G.Corcella {\it et. al.}
	 Jour. High Ener. Physics. {\bf 01} (2001)010.
	 
	\bibitem{star_white}  J. Adams {\it et al.} (STAR Collaboration)
	Nucl. Phys. A {\bf 757}, 102 (2005).
	
	\bibitem{heinz} U. W. Heinz and M. Jacob, arXiv:nucl-th/0002042 [nucl-th].
	

	
	

\bibitem{Larry_saturation} 
L. McLerran, J. Dunlop, D. Morrison, R. Venugopalan, 
Nucl. Phys. A {\bf 854} (2011) 1.
		
		\bibitem{Larry1} E. Iancu, K. Itakura, L. McLerran, Nucl. Phys. A {\bf 708} (2002) 327.
		
		\bibitem{Larry2} E. Iancu, L. McLerran, Phys. Lett. B {\bf 510} (2001) 145.
		

		
	 \bibitem{cgc1} Jamal Jalilian-Marian 
	Phys. Rev. {\bf C 70}, 027902 (2004).
	\bibitem{cgc2} D. Kharzeev and E. Levin, Phys. Lett. B {\bf 523}, 79
	(2001). 
	
	\bibitem{cgc3}	
	K. Dusling and R. Venugopalan, Phys. Rev. Lett. {\bf 108} (2012) 262001.
	
	\bibitem{duraes} F.O. Duraes, A.V. Giannini, V.p. Goucalves and F.S. Navarra, Phys. Rev. C 94, 024917
	(2016).
	\bibitem{lappi} T. Lappi, H. Mantysaari, Nucl. Phys. A 926, 186 (2014).
	
	\bibitem{rezaeian} A. H. Rezaeian, Phys. Lett. B 727, 218 (2013).
	
	\bibitem{ZEUS} M. Derrick {\it et al.} (ZEUS Collaboration) Phys. Lett. B {\bf 345} (1995) 576.
		
\bibitem{H1} C. Adlof {\it et al.} (H1 Collaboration), Phys. Lett. B {\bf 520} (2001) 183.
		
		
	
		\bibitem{benecke} J. Benecke, T. T. Chou, C. N. Yang and E. Yen, Phys.
	Rev. \textbf{188}, 2159 (1969).
	
	\bibitem{raha} R. Beckmann, S. Raha, N. Stelte, R.M. Weiner,
	Phys. Lett. {\bf B105}, 411 (1981).
	
	\bibitem{gelis} F. Gelis, A.N. Stasto, R. Venugopalan, 
	Eur. Phys. J. C \textbf{48}, 489 (2006).
	
	
	\bibitem{npart1} B. Alver {\it et al.} (PHOBOS Collaboration)
	Phys. Rev. C {\bf 94}, 024903 (2016).
	
	\bibitem{npart2} G. Torrieri, EPJ Web of Conferences, 04002 (2011).
	
	\bibitem{ALICE_XeXe} S. Acharya {\it et al.} (ALICE Collaboration) 	arXiv:1805.04432 [nucl-ex].
	
	\bibitem{ALICE_PbPb}  {J.~Adam \it et al.} (ALICE Collaboration) Phys. Lett B
	  {\bf 772}, 567 (2017).
	
	\bibitem{Larry3} L. D. McLerran, Raju Venugopalan, 
Phys. Rev. D {\bf 49} (1994) 2233.

	\bibitem{UrQMDapply1} 
	M. Bleicher {\it et al.,}, Phys. Lett. {\bf B 435}, 9 (1998). 
	
	 \bibitem{UrQMDapply2}
	M. Bleicher, S. Jeon, V. Koch, Phys. Rev. C {\bf 62}, 061902(R) (2000) 

	\bibitem{Bleichersus}
	S. Haussler, H. Stocker and M. Bleicher, Phys. Rev. C {\bf  73},
	021901(R) (2006).
	
	\bibitem{Arghya}
	A. Chatterjee, S. Chatterjee, T. K. Nayak, N. R. Sahoo, J. Phys. G:
	Nucl. Phys. J. Phys. {\bf 43}, 125103 (2016)
	
	\bibitem{ampt1} Z.-W. Lin {\it et al.}  Phys. Rev. C {\bf 72}, 064901 (2005).
	  
	\bibitem{ampt2} Z.-W. Lin {\it et al.}, Phys. Rev. C {\bf 64}, 011902 (2001). 
	
	\bibitem{ampt3}
	B. Zhang {\it et al.}, Phys. Rev. C {\bf 61},  067901 (2000).
	\bibitem{EPOS-param} A.G. Knospe {\it et al.} Phys. Rev. C {\bf 93}, 014911 (2016).
	\bibitem{EPOS3} K. Werner {\it et al.}, Phys. Rev. C {\bf 89}, 064903 (2014).
	
	\bibitem{EPOS4} K. Werner, Phys. Rev. Lett. {\bf 98}, 152301 (2007).
	\bibitem{EPOS1}
	K. Werner {\it et al.}, Phys. Rev. C {\bf 82}, 044904 (2010). 
	
	\bibitem{EPOS5} M.~Nahrgang {\it et al.}, Phys. Rev. C {\bf 90}, 024907 (2014).
	
	\bibitem{thermi} M. Chojnacki, A. Kisiel, W. Florkowski and
	  W. Broniowski, Comput. Phys. Commun. {\bf 183} (2012) 746.
	
	
	
\bibitem{pp_900_2360}
	K.~Aamodt {\it et al.} [ALICE Collaboration],
	Eur.\ Phys.\ J. C {\bf 68} (2010) 89. 	 
		 
    \bibitem{pp_2760_7_8}
	J.~Adam {\it et al.} [ALICE Collaboration],
	Eur.\ Phys.\ J. C {\bf  77} (2017) no.1,  33. 
	
	
	
		 \bibitem{pp_13}  J.~Adam {\it et al.} [ALICE Collaboration],
	Phys.\ Lett. B {\bf 753} (2016) 319. 
		 
	\bibitem{pp_630_1800}
	F.~Abe {\it et al.} [CDF Collaboration],
	Phys.\ Rev. D {\bf  41} (1990) 2330. 
	
    \bibitem{UA5} K. Alpgard {\it et al.} [UA5 Collaboration]
    Phys. Lett.B {\bf 112} (1982) 183,  G. J. Alner {\it et al.} [UA5 Collaboration]
    Phys. Rept.{\bf 154} (1987) 247.
    
		\bibitem{systematic_star} 
	B.~I.~Abelev {\it et al.} [STAR Collaboration],
	Phys.\ Rev. C {\bf 79} (2009) 034909.
	
	\bibitem{systematic_phobos}B.~Alver {\it et al.} [PHOBOS Collaboration],
	Phys.\ Rev.\ Lett.\  {\bf 102} (2009) 142301.
		  
	\bibitem{eta_dist_Au_62}
	B.~B.~Back {\it et al.} [PHOBOS Collaboration],
	Phys. Rev. C {\bf  74} (2006) 021901.
	
	\bibitem{eta_dist_Cu}
	 B.~Alver {\it et al.} [PHOBOS Collaboration],
	 Phys.\ Rev. C {\bf  83} (2011) 024913
		  
	\bibitem{eta_Au_130}
	B.~B.~Back {\it et al.} [PHOBOS Collaboration],
	Phys.\ Rev.\ Lett.\  {\bf 87} (2001) 102303
	
	\bibitem{eta_200_19}
	B.~B.~Back {\it et al.} [PHOBOS Collaboration],
	Phys.\ Rev.\ Lett.\  {\bf 91} (2003) 052303
		  
	\bibitem{eta_dist_2760}
	E.~Abbas {\it et al.} [ALICE Collaboration],
	Phys.\ Lett. B {\bf 726} (2013) 610.
	
	\bibitem{Pb_5020_eta}
	J.~Adam {\it et al.} [ALICE Collaboration],
	Phys. Lett. {\bf B 772} (2017) 567.
		  
	\bibitem{mul_dens_PbPb5020}
	J.~Adam {\it et al.} [ALICE Collaboration],
	Phys.\ Rev.\ Lett.\  {\bf 116} (2016) 222302.
	
	\bibitem{eta_dAu}
	B.~B.~Back {\it et al.} [PHOBOS Collaboration],
	Phys.\ Rev. C {\bf  72} (2005) 031901 
	
	\bibitem{pPb_eta_dist}
	G.~Aad {\it et al.} [ATLAS Collaboration],
	Eur.\ Phys.\ J. C {\bf  76} (2016) no.4,  199 
	
    \bibitem{eta_pPbs}
	B.~Abelev {\it et al.} [ALICE Collaboration],
	Phys.\ Rev.\ Lett.\  {\bf 110} (2013) no.3,  032301 
	
	\bibitem{Ncoll_pPb}
	J.~Adam {\it et al.} [ALICE Collaboration],
	Phys.\ Rev. C {\bf 91} (2015) no.6,  064905. 
	
	\bibitem{ROOT}
	R. Brun and F. Rademakers, Nucl. Inst. and Meth. in Phys. Res. A {\bf 389} (1997) 81.
	
	
	\bibitem{cgc4}	
	A. Krasnitz and R. Venugopalan, 
	Nucl. Phys. A {\bf 698} (2002) 209. 
	 
	
	\bibitem{wolschin1} B. Kellers, G. Wolschin, Prog. Theor. Exp. Phys. {\bf 2019} (2019) 053D03.

	
\bibitem{Goncalves:2019uod}
K.~J.~Gonçalves, A.~V.~Giannini, D.~D.~Chinellato and G.~Torrieri,
Phys. Rev. C \textbf{100} (2019) 054901.




	  
	  
\bibitem{Sahoo:2018osl}
P.~Sahoo, P.~Pareek, S.~K.~Tiwari and R.~Sahoo,
Phys. Rev. C \textbf{99} (2019) 044906.

	  
	  
	  	\bibitem{alice-nch} J.~Adam \textit{et al.} [ALICE Collaboration], `
	Phys. Lett. B \textbf{754} (2016), 373. 
	
	\bibitem{phobos-nch} B.~B.~Back \textit{et al.} [PHOBOS], 
	Phys. Rev. C \textbf{74} (2006), 021902. 
	  
		
		
		
		
	

		 

		  
	
		  
	\bibitem{CMS_pp}
	 V. Khachatryan {\it et al.}[CMS Collaboration]
	 Jour. High Ene. Phys. {\bf 1002} (2010) 041.
		  
	
	
	\bibitem{wolschin} G. Wolschin, Phys. Rev. C {\bf 91}, 014905 (2015).
	
	
		  
    
		  
		 
    
	
		 
	
	
	\bibitem{parton_dist}
	L. A. Harland-Lang, A. D. Martin, P. Motylinski, and R.S. Thorne, 
	Eur.Phys.J. C {\bf75} (2015) no.5, 204
		 
	
	
	
	

	
	  
	  \bibitem{he-lhc} A. Abada \textit{et al.} [FCC Collaboration], Eur. Phys. J. Special Topics 
	  {\bf 228} (2019) 1109.
	  
	  \bibitem{fcc} A. Dainese \textit{et al.} PoS HardProbes2018 (2019) 005 (e-Print: 1901.10952 [hep-ph]).
 	  
 	  	\bibitem{EKRT2000} 
	K. J. Eskola, K. Kajantie, P. V. Ruuskanen, and K.Tuominen, Nucl. Phys. {\bf B 570}, 379 (2000).
	
	\bibitem{Eskola:2015uda} 
	  K.~J.~Eskola, H.~Niemi and R.~Paatelainen,
	Nucl. and Part. Phys. Proc., {\bf 276-278} (2016) 161.
 	  
	\bibitem{alba1} J.L. Albacete, C. Marquet, 
	Progress in Particle and Nuclear Physics 76, 1 (2014).
	
	\bibitem{alba2} J. L. Albacete, A. Dumitru, Y. Nara,
	J. Phys.: Conf. Ser. {\bf 316}, 012011 (2011).

\bibitem{bass_trento}
	J. Scott Moreland, Jonah E. Bernhard, and Steffen A. Bass, Phys. Rev. C {\bf 92} (2015) 011901.
	
	\bibitem{Shen:2014vra}
	C.~Shen {\it et al.}
	Comput.\ Phys.\ Commun.\  {\bf 199} (2016) 61.

	



\end{thebibliography}
\end{document}